%% file: sage.tex
\documentclass[iop]{emulateapj}
\usepackage{natbib}
\usepackage{graphicx}
\shorttitle{Semi-Analytic Galaxy Evolution (SAGE)}
\shortauthors{Croton et al.}

\begin{document}

\title{Semi-Analytic Galaxy Evolution (SAGE): Model Calibration and Basic Results}

\author{Darren J.~Croton,\altaffilmark{1}
Adam R.~H.~Stevens,\altaffilmark{1}
Chiara Tonini,\altaffilmark{2,1}
Thibault Garel,\altaffilmark{3,1,4} 
Maksym Bernyk,\altaffilmark{1} 
Antonio Bibiano,\altaffilmark{1} 
Luke Hodkinson,\altaffilmark{1} 
Simon J.~Mutch,\altaffilmark{2,1} 
Gregory B.~Poole,\altaffilmark{2,1}
and Genevieve M.~Shattow\altaffilmark{1}}

\affil{$^1$Centre for Astrophysics \& Supercomputing, Swinburne University of Technology, PO Box 218, Hawthorn, Victoria 3122, Australia}
\affil{$^2$School of Physics, University of Melbourne, Parkville, Victoria 3010, Australia}
\affil{$^3$Centre de Recherche Astrophysique de Lyon, Universit\'{e} de Lyon, Universit\'{e} Lyon 1, CNRS, Observatoire de Lyon, 9 avenue Charles Andr\'{e}, 69561 Saint-Genis Laval Cedex, France}
\affil{$^4$Australian Research Council Super Science Fellow}

\begin{abstract}
This paper describes a new publicly available codebase for modelling galaxy formation in a cosmological context, the ``Semi-Analytic Galaxy Evolution'' model, or \textsc{sage} for short.\footnote{https://github.com/darrencroton/sage} \textsc{sage} is a significant update to that used in \citet{Croton2006} and has been rebuilt to be modular and customisable. The model will run on any $N$-body simulation whose trees are organised in a supported format and contain a minimum set of basic halo properties. In this work we present the baryonic prescriptions implemented in \textsc{sage} to describe the formation and evolution of galaxies, and their calibration for three $N$-body simulations: Millennium, Bolshoi, and GiggleZ. Updated physics include: gas accretion, ejection due to feedback, and reincorporation via the galactic fountain; a new gas cooling--radio mode active galactic nucleus (AGN) heating cycle; AGN feedback in the quasar mode; a new treatment of gas in satellite galaxies; and galaxy mergers, disruption, and the build-up of intra-cluster stars. Throughout, we show the results of a common default parameterization on each simulation, with a focus on the local galaxy population.
\end{abstract}


\maketitle

\input{section01}

\input{section02}

\input{section03}
\input{section04}

\input{section05}
\input{section06}
\input{section07}

\input{section08}

\input{section09}
\input{section10}
\input{section11}
\input{section12}

\input{section14}

\section*{Acknowledgements}

DC acknowledges receipt of a QEII Fellowship from the Australian Research Council. 
TG acknowledges support from an Australian Research Council Super Science Fellowship, and is grateful to the LABEX Lyon Institute of Origins (ANR-10-LABX-0066) at the Universit\'e de Lyon for its financial support within the program ``Investissements d'Avenir'' (ANR-11-IDEX-0007) of the French government, operated by the National Research Agency (ANR).

The Millennium Simulation was carried out by the Virgo Supercomputing Consortium at the Computing Centre of the Max Plank Society in Garching. It is publicly available at http://www.mpa-garching.mpg.de/Millennium/.
The Bolshoi Simulation was carried out by A.~Klypin, J.~Primack and S.~Gottloeber at the NASA Ames Research Centre. The simulation and data products can be found at http://astronomy.nmsu.edu/aklypin/Bolshoi/.
The GiggleZ simulations were conducted by G.~Poole on the Green supercomputer at Swinburne University with subsequent analysis conducted with Swinburne's gSTAR cluster.

Some data used in this work were generated using Swinburne University's Theoretical Astrophysical Observatory (TAO). TAO is part of the Australian All-Sky Virtual Observatory (ASVO) and is freely accessible at https://tao.asvo.org.au.  

ARHS thanks Toby Brown for sourcing the relevant SDSS data used in Fig.~\ref{fig:quiescent} and discussion on the quiescent fraction.  Funding for SDSS-III has been provided by the Alfred P. Sloan Foundation, the Participating Institutions, the National Science Foundation, and the U.S. Department of Energy Office of Science. The SDSS-III web site is http://www.sdss3.org/.

Finally, the authors would like to thank the anonymous referee, whose careful reading of this paper resulted in many valuable improvements.

\bibliographystyle{mn2e}
\bibliography{./sage}

\label{lastpage}

\end{document}

%% file: section01.tex
\section{Introduction}
\label{sec:intro}

Developing a complete theory of galaxy evolution is a formidable task.  Without the ability to construct real universes in a laboratory, we are left to test ideas through conducting supercomputer simulations and comparing their results against what we observe.  Arguably the most thorough way of doing this currently is through cosmological hydrodynamic simulations \citep[e.g. \citealt*{Carlberg1990};][]{Dubois2014c,Vogelsberger2014a,Khandai2015,Schaye2015}, where the physics of baryons and the dark sector are self-consistently considered.  However, one can successfully reproduce the large-scale structure and formation sites of galaxies inside halos with pure $N$-body simulations (e.g.~\citealt{Davis1985,Springel2005,Kim2011}; \citealt*{Klypin2011}; \citealt{Skillman2014,Poole2015}), into which galaxies can be later added in post-processing. Such simplicity can be significantly advantageous.

By foregoing a simulation of costly hydrodynamic processes, simulators can invest their computational resources in increasing the number of particles. Structures on both smaller and larger scales are then resolved through improved particle mass resolution and by simulating larger volumes, respectively. Larger volumes also lead to less biased sampling and more halos overall, which in turn allow for better statistical significance.  For example, the main run of the hydrodynamic EAGLE simulations \citep{Schaye2015} used $1504^3$ dark matter particles, but required approximately the same number of floating point operations as the main run of the pure $N$-body Dark Sky simulations \citep{Skillman2014} with $10240^3$ particles.  The addition of hydrodynamics comes at the cost of approximately two orders of magnitude in particle number.

Semi-analytic models of galaxy evolution take advantage of the relative computational efficiency of $N$-body simulations by adding the bound baryons to a simulation as a post-processing step.  By using information about the gravitationally bound halos, such as their mass, size, spin, substructure, and merger history, the properties of galaxies hosted within these structures can be inferred through differential equations that describe the relevant physics macroscopically.  The runtime of such a model is orders of magnitude less than the $N$-body simulation itself.  Hence, after that initial investment one can explore large regions of the parameter space that underly the baryonic physics by running the model many times.  

In the original implementation of the semi-analytic method \citep{White1991}, halos were drawn statistically from the Press-Schechter formalism for gravitational collapse \citep[\citealt{Press1974}; later extended by][]{Bond1991,Bower1991}, and the analytic baryonic physics was based on the milestone theory of \citet{White1978}.  \citet{Kauffmann1993} introduced the use of merger trees for a model, where individual systems could then be tracked across time statistically within Press-Schechter theory.  Merger trees produced from an $N$-body simulation should always be more representative of the real Universe, and \citet{Kauffmann1999} were the first to utilise this technique.  Now, it is standard practice for semi-analytic models to be coupled to an $N$-body simulation \citep[e.g.][]{Hatton2003,Bower2006,Croton2006,Benson2012,Henriques2014}.  For a more in-depth look at the history of semi-analytic models, see \citet{Baugh2006}.

In this paper, we present our new semi-analytic model, \textsc{sage} (Semi-Analytic Galaxy Evolution), which updates the work of \citet[][hereafter C06]{Croton2006}.  The new model revamps many prescriptions for the treatment of baryons, including the suppression of cooling within halos from active galactic nuclei (AGN) feedback, reincorporation of ejected gas, and the stripping of gas from satellite systems.  While semi-analytic models have historically been designed and calibrated for a single $N$-body simulation, \textsc{sage} is designed to be run on any simulation, so long as the merger trees are provided in an appropriate format.  We show \textsc{sage}'s performance on three cosmological $N$-body simulations; namely the Millennium \citep{Springel2005}, Bolshoi \citep{Klypin2011}, and GiggleZ \citep{Poole2015} simulations, all of which follow the standard $\Lambda$CDM cosmological paradigm. That said, there is nothing to stop \textsc{sage} being run on simulations that explore alternate gravity models, or different dark matter candidates, for example.

Along with those from other semi-analytic models, \textsc{sage} galaxy catalogues built on various $N$-body simulations are available through the Theoretical Astrophysical Observatory \citep{Bernyk2014}.\footnote{https://tao.asvo.org.au}  The codebase of \textsc{sage} is also publicly available,\footnote{https://github.com/darrencroton/sage} allowing the community to build further models, or modify those described here.  The repository includes an \textsc{ipython} notebook for conducting simple analysis with \textsc{sage} output, specifically showing how we produced some of the figures presented in this paper.\footnote{This can be viewed directly on or downloaded from GitHub at the above link. Once downloaded, the notebook will function without needing to run \textsc{sage} first (it will fetch output).}

We lay out the sections of this paper in the following manner.  The $N$-body simulations used as input for \textsc{sage} are summarised in Section \ref{sec:sims}.  We provide an overview of \textsc{sage} in Section \ref{sec:model}, covering which components of the model have been upgraded from that of C06.  Sections \ref{sec:infall}-\ref{sec:bursts} then describe the physics of the model in more detail, covering: gas infall in halos (\ref{sec:infall}); the role of reionization (\ref{sec:reion}); cooling of gas from the hot halo (\ref{sec:cooling}); the consideration of cold gas, star formation, and metal enrichment in galactic disks (\ref{sec:sf}); the role of supernova feedback (\ref{sec:sn}); the growth of black holes and their associated AGN feedback (\ref{sec:agn}); dealing with mergers and intra-cluster stars (\ref{sec:merging}); disk instabilities (\ref{sec:instability}); and starbursts (\ref{sec:bursts}). We finish with some discussion and concluding remarks in Section \ref{sec:discussion}.

Throughout, we present all results assuming $h=0.73$, based on the cosmology used for the Millennium simulation, the primary simulation used for calibrating the model. Where relevant, we also use a \citet{Chabrier2003} initial mass function to produce stellar masses.

%% file: section02.tex
\section{N-body simulations}
\label{sec:sims}

In this paper, we present the performance of \textsc{sage} on three cosmological simulations.  Each simulation not only varies in terms of its cosmological parameters, having used the best available measurements at their various times of being run, but also in terms of the codes and pipelines used to generate the final data products.  Even for simulations with identical initial conditions and cosmological parameters, the use of different codes for either running or post-processing cosmological simulations can lead to non-trivial differences in their results.  \citet{Knebe2011,Knebe2013} assessed how and why the choice of (sub)halo finder can affect results, while complimentary studies investigated how the choice of merger tree code can change the derived structure formation histories and the consequences this has for semi-analytic models \citep[][respectively]{Srisawat2013,Lee2014}.  Understanding these non-trivial differences is key to understanding the theoretical and numerical uncertainties associated with semi-analytic models, although we do not attempt to provide a detailed analysis of such effects here.  We describe each simulation below and summarise their properties in Table \ref{tab:sims}.

\begin{table*}
	\centering
	\begin{tabular}{l r r r r r r l l l}\hline\hline
		Simulation & $N_{\rm part}$ & $M_{\rm part}h$ & $l_{\rm box}h$  & $\Omega_M$ & $\sigma_8$ & Code & Subhalo finder & Tree constructor\\
		 & & (M$_{\bigodot}$) & (cMpc)\\\hline\hline
		Millennium & $2160^3$ & $8.60 \times 10^8$ & 500 & 0.250 & 0.900 & 
		\textsc{gadget-2} & \textsc{subfind} & \textsc{l-halotree}\\
		Bolshoi & $2048^3$ & $1.35 \times 10^8$ & 250 & 0.270 & 0.820 & \textsc{art} & \textsc{rockstar} & \textsc{consistent-trees}\\
		GiggleZ-MR & $520^3$ & $9.50 \times 10^8$ & 125 & 0.273 & 0.812 & \textsc{gadget-2} & \textsc{subfind} & Poole et al.~(in prep.)\\\hline\hline
	\end{tabular}
	\caption{Details of the $N$-body simulations used in the analysis of \textsc{sage} in this paper.  The columns provide particle number, $N_{\rm part}$; particle mass, $M_{\rm part}$; periodic box length, $l_{\rm box}$, in comoving units; contribution of matter to the average universal energy density at the present epoch, $\Omega_M$ (for which the equivalent for dark energy is $\Omega_{\Lambda} = 1-\Omega_M$); the redshift-zero extrapolation of the root-mean-square linear mass fluctuation within a sphere of radius $8h^{-1}$ Mpc, $\sigma_8$; the code the simulation was run with; the code used to identify subhalos; and the code used to build the merger trees.}
	\label{tab:sims}
\end{table*}

\subsection{The Millennium simulation}
\label{sec:millennium}

The Millennium simulation \citep{Springel2005} significantly upped the ante in cosmological simulations, boasting unrivalled detail for its time.  It has since been the focus of a plethora of scientific studies.\footnote{See http://www.mpa-garching.mpg.de/millennium/ for an up-to-date list.}  Ten years on from its completion, the simulation remains a benchmark, and continues to be used for science.  Of the simulations used in this paper, it remains the most balanced in terms of (cosmologically representative) size and resolution (cf.~Table \ref{tab:sims}).

Millennium was run using the popular \textsc{gadget-2} code \citep{Springel2005Gadget2}. The cosmological parameters followed those from a combined analysis of \emph{WMAP}1 (Wilkinson Microwave Anisotropy Probe, first year) data \citep{Spergel2003} and the 2-degree Field Galaxy Redshift Survey \citep{Colless2001}.  Arguably, the biggest weakness of Millennium is its dated cosmological parameters, which now differ significantly from the best-fitting values which are more precisely measured \citep[for the latest observational results, see][]{Planck2015XIII}.

Structure and subhalos were identified in the Millennium simulation with the \textsc{subfind} algorithm \citep{Springel2001}.  Parent halos are found through a friends-of-friends procedure, while subhalos are defined as having at least 20 gravitationally bound particles by the halo finder.  The merger trees which feed \textsc{sage} were constructed with the \textsc{l-halotree} code \citep[described in the supplementary information of][]{Springel2005}.

\subsection{The Bolshoi simulation}
\label{sec:bolshoi}

Bolshoi \citep{Klypin2011} was run using the adaptive-mesh-refinement code \textsc{art} \citep*[Adaptive Refinement Tree,][]{Kravtsov1997}.  The chosen cosmological parameters were very close to those of the \emph{WMAP}7 data \citep{Jarosik2011}, while maintaining consistency with \emph{WMAP}5 (\citealt{Dunkley2009}; also see \citealt{Komatsu2009}).  When compared with \emph{WMAP}1, the data from these \emph{WMAP} releases describe a universe with a greater average matter density, that presently expands more slowly, with smaller mass fluctuations.  While smaller in box size, Bolshoi complements the results from Millennium due to its higher mass resolution, allowing us to probe the low-mass end of the mass function in more detail.

Subhalos in Bolshoi were identified with the \textsc{rockstar} algorithm \citep*{Behroozi2013a},\footnote{Robust Overdensity Calculation using K-Space Topologically Adaptive Refinement, available at https://bitbucket.org/gfcstanford/rockstar} which builds a hierarchy of friends-of-friends subgroups and utilises one temporal and six phase-space quantities to determine which particles constitute those subhalos.  Merger trees were subsequently constructed using \textsc{consistent-trees} \citep{Behroozi2013b}.\footnote{https://bitbucket.org/pbehroozi/consistent-trees}

\subsection{The GiggleZ simulation suite}
\label{sec:gigglez}

The Gigaparsec WiggleZ simulation suite \citep[GiggleZ,][]{Poole2015} was run as a theoretical counterpart to the WiggleZ Dark Energy Survey \citep{Drinkwater2010}.  Each simulation was performed using \textsc{gadget-2}, with cosmological parameters based on data from \emph{WMAP}5, baryonic acoustic oscillations, and supernovae \citep{Komatsu2009}.  While the main box of GiggleZ boasts shear size (1$h^{-1}$ comoving Gpc on a side), due to its lower mass resolution we instead assess the complimentary GiggleZ-MR simulation for this paper. GiggleZ-MR was run in a smaller box of side length 125$h^{-1}$ comoving Mpc but with a particle mass much closer to Millennium.

GiggleZ subhalos were identified with \textsc{subfind}. Trees were built following the method in Poole et al.~(in preparation). This approach repairs pathological defects in merger trees introduced by the halo-finding process (e.g.~over linking, or the disappearance of halos during pericentric passages) through a process of forward and backward matching which scans both ways over multiple snapshots.

\subsection{Halo merger tree structure and required properties}
\label{sec:data}

\textsc{sage} is modular enough that it should run on any halo merger tree that is structured in a supported format and contains a minimum set of information per halo. The initial public release of \textsc{sage} requires halo measurements for:
\begin{itemize}
\item $M_{\rm vir}$, the halo virial mass, and `Len', the number of particles in the (sub)halo;
\item $V_{\rm vir}$ and $R_{\rm vir}$, measured using the standard virial relations;
\item $V_{\rm max}$, a fit to the maximum circular velocity of the halo; and
\item the cartesian position, velocity, and spin vector of each halo.
\end{itemize}
Additional input properties may be required depending on the version of \textsc{sage} being used. This can be checked in core\_simulation.h, where the tree file structure is defined, and also directly in each relevant core or science module.

In terms of tree structure, \textsc{sage} assumes the halo trees are organised in depth-first order, as described in the supplementary information of \citet[][see, in particular, their fig.~5]{Springel2005}. This is the same output format used for the trees of the Millennium simulation (Section \ref{sec:millennium}), which were constructed with the \textsc{gadget-2}/\textsc{l-halotree} codebase. \textsc{l-halotree} produces additional output properties required by \textsc{sage} that enable the membership and history of each halo to be determined:
\begin{itemize}
\item `FirstProgenitor', `NextProgenitor', `FirstHaloInFOFgroup', and `NextHaloInFOFgroup'.
\end{itemize}
Given the modular nature of \textsc{sage}, other tree formats, such as those produced by \textsc{rockstar}/\textsc{consistent-trees}, may be supported in the future.

%% file: section03.tex
\section{An overview of \textsc{sage}}
\label{sec:model}

The new \textsc{sage} model of galaxy formation is an updated version of that described in C06. One of the most important aspects of this update is that the code has been cleaned, significantly optimised, generalised to run on a wide variety of simulations, and ready to be used by the wider community. Key changes to the physics that distinguish \textsc{sage} from the C06 model are summarised below. A more detailed description of each model component is then given in the subsequent sections, where we also explore their consequences on the galaxy population and its evolution. 

\begin{itemize}

\item \emph{Gas cooling and AGN heating}: Cooling and heating of halo gas is now more directly coupled in the new \textsc{sage} model. We introduce a heating radius from radio mode AGN feedback, interior to which gas will never cool. This radius can only move outward, thus retaining the memory of previous heating episodes. The cooling rate is then calculated using only the hot gas between the cooling and heating radii.\\

\item \emph{Quasar mode feedback}: We have added feedback from the quasar mode, the dominant growth channel of the black hole, triggered by mergers or disk instabilities. This feedback is most effective at removing disk (and sometimes halo) gas at high redshift in gas rich galaxies.\\

\item \emph{Ejected gas reincorporation}: Gas ejected from the halo is now reincorporated according to the dynamical time of the halo modulated by $V_{\rm vir}/V_{\rm crit}$, with $V_{\rm crit}$ a parameter set at the galaxy group halo mass scale. Previously, reincorporation was dependent on the dynamical time alone.\\

\item \emph{Satellite galaxies}: Hot gas is no longer instantaneously removed from the satellite/subhalo system upon infall, but stripped in proportion to the subhalo dark matter mass stripping rate. Satellites are treated in the same way as central galaxies for the longest time possible (e.g.~we allow cooling in subhalos).\\
 
\item \emph{Mergers and intra-cluster stars}: At the moment of infall, an expected satellite merger time is calculated. The satellite is then tracked until its baryon-to-subhalo mass falls below a critical threshold (taken at $\sim 1$). At this point the current survival time is compared to the expected merger time. If the subhalo has survived longer than expected we say it is more resistant to disruption and the satellite is merged with the central in the usual way. Otherwise the satellite is disrupted and its stars are added to a new intra-``cluster'' mass component. As a consequence, \textsc{sage} no longer produces satellite galaxies lacking a subhalo, the so-called orphan population.
\end{itemize}

\subsection{Model calibration}
A summary of the primary \textsc{sage} parameters, along with the choices that define our fiducial galaxy model, are given in Table \ref{tab:params} for quick reference. These parameters were manually selected to simultaneously perform well across all three simulation sets, with a slightly higher emphasis on Millennium. All figures and results in this paper are calculated using a model with these parameter choices.  Note that, due to the order-of-magnitude higher resolution of Bolshoi, we found it necessary to lower its baryon fraction parameter (only) from 0.17 to 0.13 to obtain comparable results to Millennium and GiggleZ-MR (see Section~\ref{sec:infall}).

\begin{table*}
	\centering
	\begin{tabular}{l l l l l c} \hline \hline
		Parameter & Description & Value & C06 value & Fixed & Section(s)\\ \hline \hline
		$f_b^{\rm (cosmic)}$ & (Cosmic) baryon fraction & 0.17, 0.13 & 0.17 & No & \ref{sec:infall}, \ref{sec:reion}\\ \hline
		$z_0$ & Redshift when H\,\textsc{ii} regions overlap & 8.0 & 8.0 & Yes & \ref{sec:reion}\\
		$z_r$ & Redshift when the intergalactic medium is fully reionized & 7.0 & 7.0 & Yes & \ref{sec:reion}\\ \hline
		$\alpha_{\rm SF}$ & Star formation efficiency & 0.05 & 0.07 & No & \ref{sec:sf}\\
		$Y$ & Yield of metals from new stars & 0.025 & 0.03 & No & \ref{sec:sf}\\
		$\mathcal{R}$ & Instantaneous recycling fraction & 0.43 & 0.30 & Yes & \ref{sec:sf}, \ref{sec:sn}\\ \hline
		$\epsilon_{\rm disc}$ & Mass-loading factor due to supernovae & 3.0 & 3.5 & No & \ref{sec:sn}\\
		$\epsilon_{\rm halo}$ & Efficiency of supernovae to unbind gas from the hot halo & 0.3 & 0.35 & No & \ref{sec:sn}\\
		$k_{\rm reinc}$ & Sets velocity scale for gas reincorporation & 0.15 & N/A & Yes & \ref{sec:sn}\\ \hline
		$\kappa_{\rm R}$ & Radio mode feedback efficiency & 0.08 & N/A & No & \ref{sec:radiomode}\\
		$\kappa_{\rm Q}$ & Quasar mode feedback efficiency & 0.005 & N/A & No & \ref{sec:quasarmode}\\ 
		$f_{\rm BH}$ & Rate of black hole growth during quasar mode & 0.015 & 0.03  & No & \ref{sec:quasarmode}\\ \hline
		$f_{\rm friction}$ & Threshold subhalo-to-baryonic mass for satellite disruption or merging & 1.0 & N/A & Yes & \ref{sec:merging}\\
		$f_{\rm major}$ & Threshold mass ratio for merger to be major & 0.3 & 0.3 & Yes & \ref{sec:merging}\\ \hline\hline
	\end{tabular}
	\caption{Fiducial \textsc{sage} parameters used throughout this text, and also compared to those from C06. The fifth column indicates whether the value was kept fixed or used in the calibration.  The last column indicates the section where the parameter is discussed.  The cosmic baryon fraction used for Millennium and GiggleZ-MR is 0.17, while 0.13 was used for Bolshoi. All other parameters are common across all three simulations. }
	\label{tab:params}
\end{table*}

\begin{figure}
	\centering
	\includegraphics[width=0.45\textwidth]{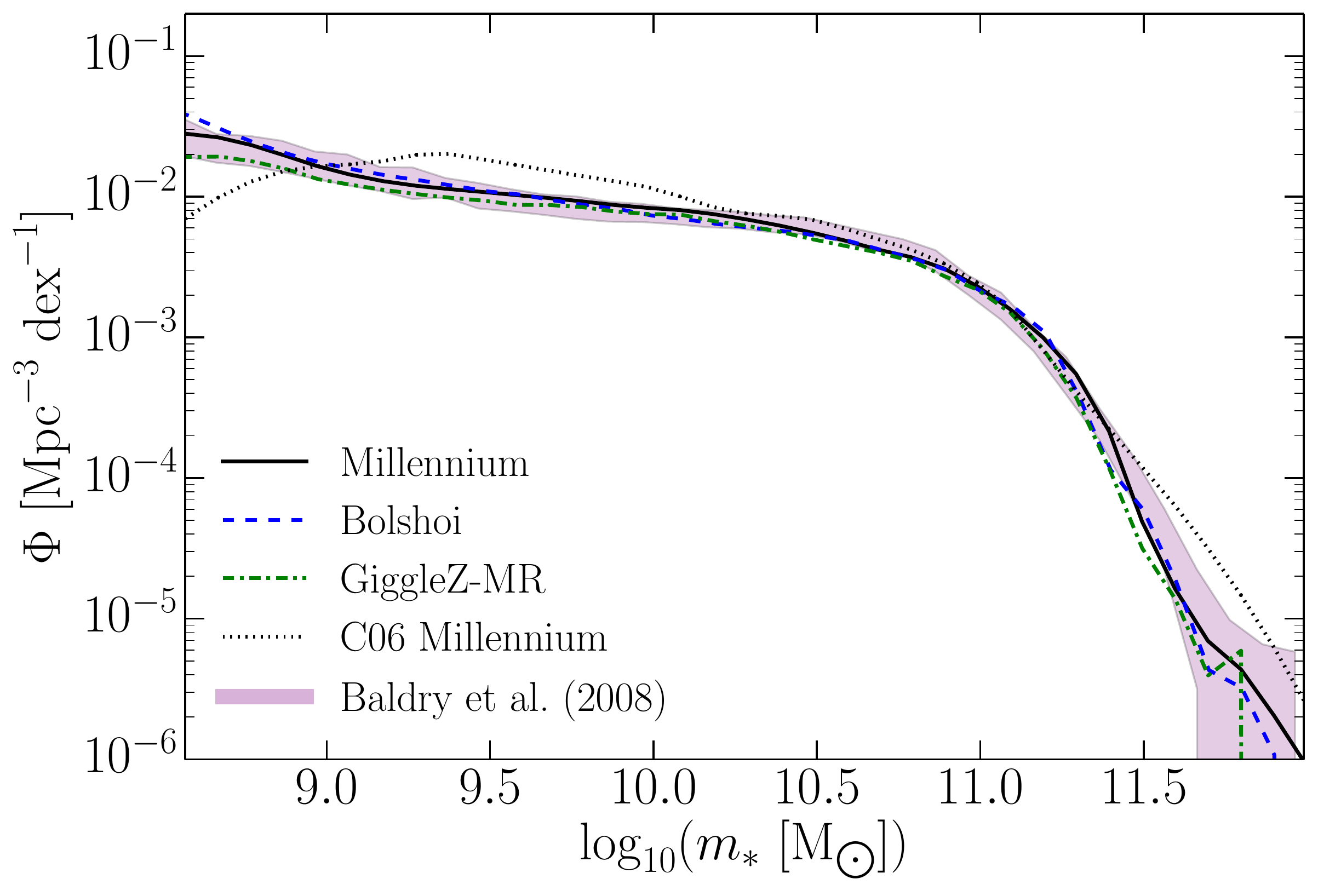}
	\caption{Stellar mass functions at $z=0$ for \textsc{sage} galaxies for each of the $N$-body simulations compared to observational data from \citet{Baldry2008} and the published C06 model, taking $h=0.73$.}
	\label{fig:smf}
\end{figure}

The primary constraint for our parameter choices was the ability to reproduce the $z=0$ stellar mass function: that is, the number density of galaxies as a function of their stellar mass, $\Phi$.  We show in Fig.~\ref{fig:smf} that \textsc{sage} produces an excellent stellar mass function for each simulation with our fiducial parameter set, tightly matching the observational uncertainty range presented by \citet*{Baldry2008}.  To conservatively account for the different simulation resolution limits, a minimum stellar mass equal to the median of galaxies in Millennium (sub)halos made of 50 particles is adopted.  We also compare the \textsc{sage} mass functions to the C06 model, but note that C06 used the luminosity function as its primary constraint instead.

We further used a set of secondary constraints for setting our fiducial parameter values.  These constraints include the star formation rate density history \citep*[][our Fig.~\ref{fig:sfrd}]{Somerville2001}, the Baryonic Tully--Fisher relation \citep*[][our Fig.~\ref{fig:btf}]{Stark2009}, the mass--metallicity relation of galaxies \citep[][our Fig.~\ref{fig:massmet}]{Tremonti2004}, and the black hole--bulge mass relation \citep*[][our Fig.~\ref{fig:bhbulge}]{Scott2013}.  We detail which parameters were allowed to be varied in the model calibration in order to meet these constraints in Table \ref{tab:params}.  All of the free parameters and several of the fixed parameters have different values to C06.

For the remainder of the paper, we walk through the different baryonic reservoirs in the model and the physics that describe how mass and energy move between them.

%% file: section04.tex
\section{Gas infall in halos}
\label{sec:infall}

Perturbations in the primordial density field lead to gravitational instability that drives mass to continually collapse into `halo'-like structures. This process is dominated by dark matter, being the main mass component of the Universe, with the baryons tending to follow rather than lead the dynamical evolution. In a cosmological $N$-body simulation, such halo growth is typically well measured by a halo finder. Hence, the total \textit{baryonic} content of each halo, and how it changes with time, can be inferred.

In the C06 model, as in \textsc{sage}, the total baryonic content of each dark matter halo at (nearly) all times is maintained at the universal fraction, $f_{\rm b}$. This requires that for each simulation time-step, the baryons in each halo grow by ${f_{\rm b}}{M_{\rm vir}}-{m_{\rm b}}$, where $m_{\rm b}$ is the total mass of baryons present in the previous time-step. This mass difference is added to the hot gas reservoir of the system and assumed to be pristine.  If halo mass were to decrease over a time-step, then ejected gas (see Section \ref{sec:sn}) or, secondarily, hot gas is removed from the system (cold gas and stars, located deep in the potential well of the halo, are always unaffected here).  However, in low-mass halos and at high redshift, little to none of the baryons are expected to be hot \citep{Birnboim2003,Keres2005}; a decrease in halo mass can then lead to a temporary increase in the baryon fraction, i.e.~above universal.

However unlike C06, \textsc{sage} considers the baryon fraction to be a \textit{free} parameter during model calibration, which sets the baseline level of baryons within the virial radius of each halo. For the Millennium and GiggleZ-MR simulations, the original value of 0.17 provides the baryonic mass required to produce a well-calibrated model. However we find that the Bolshoi simulation, with its order-of-magnitude higher mass resolution, needs a lower baseline fraction of baryons (assuming the same remaining model parameters), $f_{\rm b}=0.13$, to obtain comparable results.

Investigating this further, it is interesting to see how the baryon fraction and simulation resolution limit together influence which halos get gas at each epoch and when this gas turns into stars. Since halos are identified earlier in Bolshoi, and given the power-law steepness of the halo mass function, a considerable fraction of baryons accumulate in low-mass halos but are delayed in their opportunity to turn into stars by the reionization prescription (see next Section). Our choice of a lower baryon fraction compensates for this accumulation by lessening the significance of such baryons once they are later able to contribute to galaxy evolution. A simple back-of-the-envelope calculation reveals that a model with $f_{\rm b}=0.17$ on Millennium results in approximately the same baryonic mass per unit volume as a model with $f_{\rm b}=0.13$ on Bolshoi.

\begin{figure}
	\centering
	\includegraphics[width=0.45\textwidth]{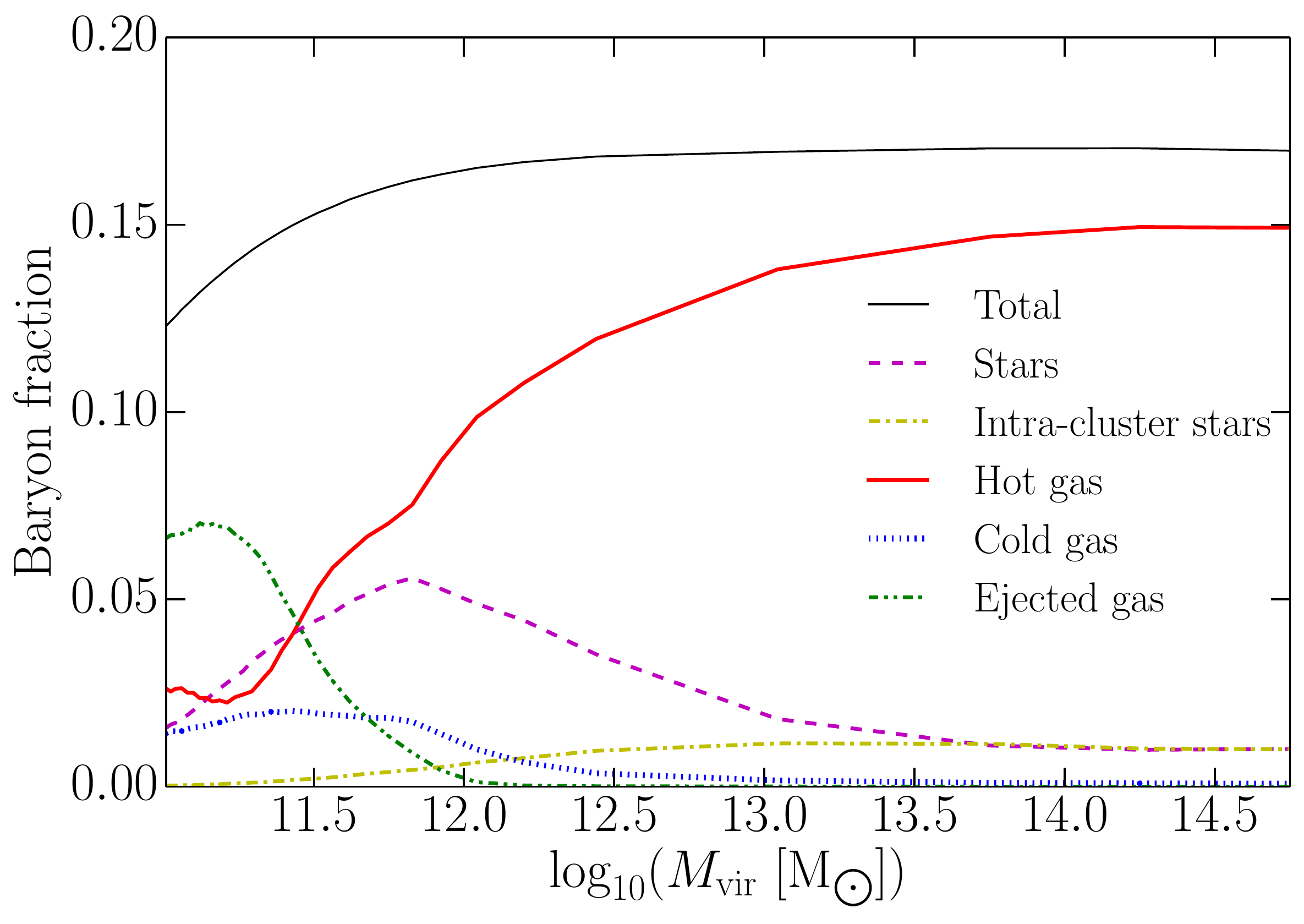}
	\caption{Fraction of baryonic content to total mass for Millennium halos at $z=0$ from \textsc{sage}, binned by virial mass.  Each curve represents the mean fraction of the labelled component.}
	\label{fig:fb}
\end{figure}

In Fig.~\ref{fig:fb} we show the baryon fraction contained within Millennium halos of a given mass at $z=0$, broken down by mass component: galaxy stars, ``intra-cluster'' stars, hot gas, cold gas, and ejected gas (these will each be defined in the subsequent Sections). One can see the dominance of hot gas in group and cluster mass systems, with stellar mass peaking in Milky Way-sized halos, and cold gas dominating at lower mass scales.

%% file: section05.tex
\section{Reionization}
\label{sec:reion}

At high redshift, the baryonic content of low-mass halos is most likely suppressed as a result of photoionization heating of the intergalactic medium (IGM) by strong feedback from the first stars. This heating acts to lower the concentration of baryons in shallow potentials \citep{Efstathiou1992}. Observationally, this can be seen in the low abundance of local dwarf galaxies relative to the prediction that results from the $\Lambda$CDM halo mass function.\footnote{This has been referred to in the literature as the ``missing satellites problem'' \citep{Moore1999,Klypin1999}, for which baryonic physical processes, including reionization heating, are capable of solving \citep[e.g.][]{Zolotov2012,Brooks2013}.}

\cite{Gnedin2000} showed using high-resolution $\rm{SLH}$-$\rm{P^3M}$ (softened Lagrangian hydrodynamics) simulations that the effect of photoionization heating can be modelled by defining a filtering mass, $M_{\rm F}$, below which the baryonic fraction in the halo is reduced relative to the universal value:
\begin{equation}
f_{\rm b}^{\rm halo}(z,M_{\rm vir}) = \frac{f_{\rm b}^{\rm cosmic}}{(1
  + 0.26 \,M_{\rm F}(z) / M_{\rm vir})^3}~.
\label{reion}
\end{equation}
Notice that the filtering mass is a function of time. In their simulations, the change was sharpest around the epoch of reionization. 

\citet*{Kravtsov2004} fitted an analytic model to approximate the behaviour of the \cite{Gnedin2000} filtering mass. They defined two parameters that characterise this transition: $z_0$, which marks the redshift where the first H\,\textsc{ii} regions overlap, and $z_r$, which marks the time when the intergalactic medium is fully reionized. The best fit values to the \citeauthor{Gnedin2000} simulation are $z_0\!=\!8$ and $z_r\!=\!7$. We adopt these in our model, leaving it identical to that used in C06.  See appendix B of \citet{Kravtsov2004} for the expression of $M_{\rm F}(z)$.

\begin{figure}
	\centering
	\includegraphics[width=0.45\textwidth]{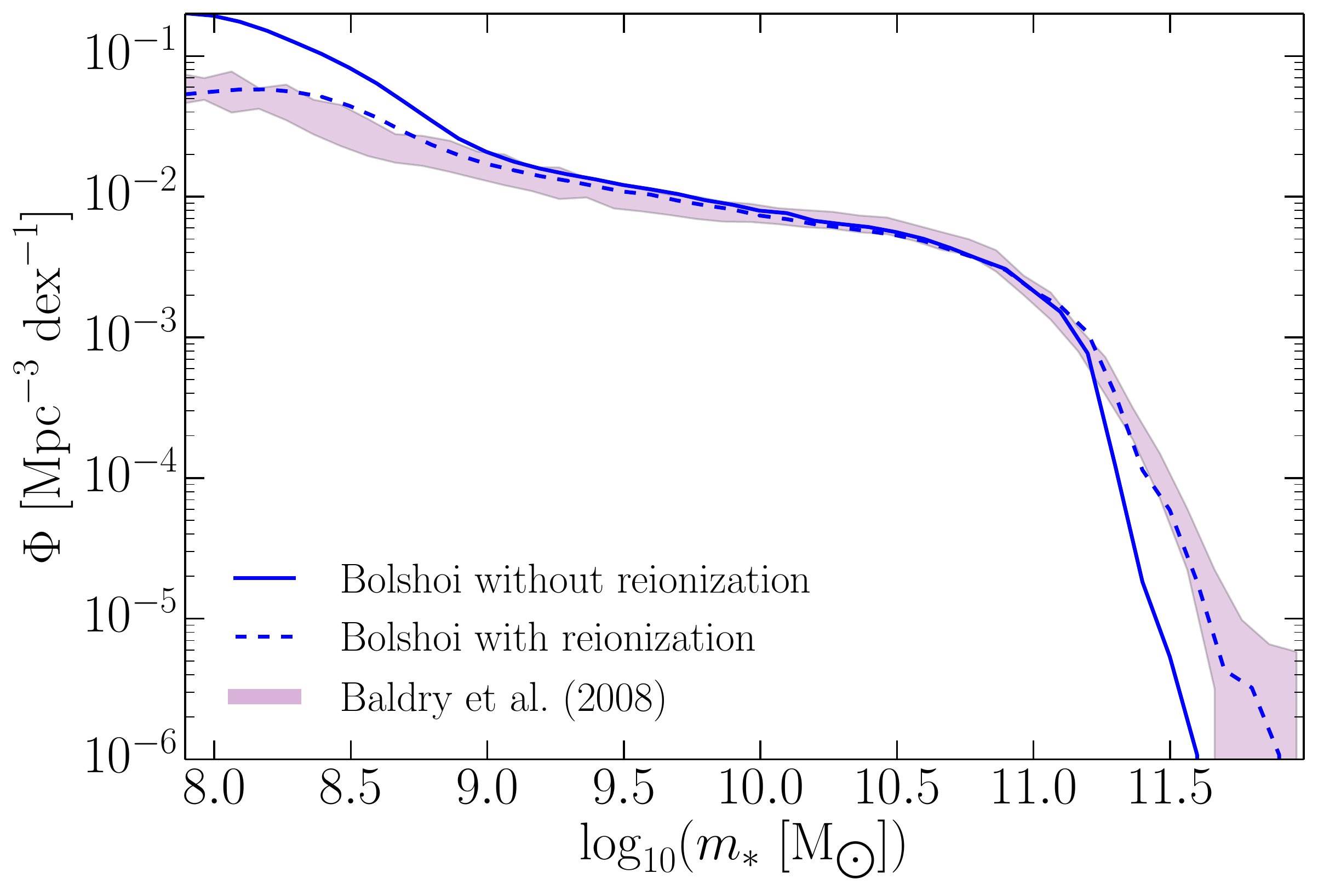}
	\caption{The effect of reionization in \textsc{sage} on the $z=0$ stellar mass function for Bolshoi.  Because of the simulation's relatively high resolution, reionization is needed to get the low-mass end of the mass function right, while also affecting the high-mass end.}
	\label{fig:reion}
\end{figure}

In Fig.~\ref{fig:reion}, we show how reionization can have an important and positive effect on both the low-mass and high-mass ends of the stellar mass function at $z=0$.  We show this for Bolshoi, using all galaxies more massive than the median stellar mass in (sub)halos made of 50 particles, which keeps us above the simulation resolution limit. Because reionization suppresses gas infall (and therefore star formation) at high redshift, satellite galaxies at $z=0$ have less stars when reionization is turned on.  Instead this gas cools at later epochs, being brought into more massive halos by the satellites, and contributes directly to the higher-mass central galaxy.  As such, these systems form more stars, and the high-mass end of the mass function kicks out.  Higher-resolution simulations have more subhalos, especially at high redshift, in which more gas can cool, and the strength of this effect is resolution-dependent (hence the effect is stronger for Bolshoi).

%% file: section06.tex
\section{The hot gas halo}
\label{sec:cooling}

The physics of infalling gas is complicated and a detailed accounting is beyond the scope of the semi-analytic methodology. Its evolution can be approximated at a level of accuracy consistent with the rest of the model by assuming that new gas added is shock heated to the virial temperature of the system as it loses its gravitational potential energy. This results in the formation of a quasistatic hot halo of baryons. However, at early times and in low-mass systems, the shock heating may be unstable, or the shock not form at all. In this case, the gas rapidly radiates away any gained energy and instead is channelled to the centre as a cold stream, the so-called `cold accretion' or `rapid cooling' mode. See Section~3.2.1 of C06 for further discussion.

In \textsc{sage}, we follow the hot-halo model described in C06. That model, based on the original work of \cite{White1991}, assumes that halo gas initially settles into an isothermal density profile at the virial temperature of the system, %
\begin{equation}
	T_{\rm vir} = \frac{1}{2}~\frac{\bar{\mu} m_p}{k}~V_{\rm vir}^2 = 35.9 \left(\frac{V_{\mathrm{vir}}}{\mathrm{km~s}^{-1}} \right)^2~\mathrm{K}~,
	\label{eq:Tvir}
\end{equation}
where $V_{\rm vir}$ is the halo virial velocity and $\bar{\mu} m_p$ is the mean particle mass, with $\bar{\mu}=0.59$ nominally. The density profile of the gas is thus
\begin{equation}
\rho_g(r) = \frac{m_{\rm hot}}{4 \pi R_{\rm vir} r^2} ~,
\label{rhog}
\end{equation}
where $m_{\rm hot}$ is the total hot gas mass in the halo which extends out to the virial radius $R_{\rm vir}$.  

To calculate the rate at which gas cools out of this distribution, the similarity solutions of \citet{Bertschinger1989} are used. We define the cooling radius, $r_{\rm cool}$, as the radius at which the gas cooling time is equal to the dynamical time of the system, $t_{\rm cool} = R_{\rm vir} / V_{\rm vir}$ (although other time-scales can be used, for example the age of the system -- see \citealt{Lu2011}). \citet{Bertschinger1989} showed that the cooling mass flux across this radius is proportional to the mass deposition rate at the centre of the halo, with the proportionality close to one.

For a parcel of gas with local density $\rho_g(r)$ and temperature $T_{\rm vir}$, the cooling time can be approximated by the ratio of its specific thermal energy to the cooling rate per unit volume,
\begin{equation}
t_{\rm cool} = \frac{3}{2} \frac{\bar{\mu}\, m_p\, k\,T_{\rm vir} }{ \rho_g(r)\, \Lambda(T_{\rm vir},Z)} ~.
\label{tcool}
\end{equation}
Here $k$ is the Boltzmann constant and $\Lambda (T,Z)$ is the cooling function, dependent on both the gas temperature and metallicity $Z$ (metal production is described below in Section~\ref{sec:sf}).

Combining Equations~\ref{rhog} and \ref{tcool}, and assuming $t_{\rm cool}$ above, allows us to find $r_{\rm cool}$ for each system. We then solve the continuity equation
\begin{equation}
\dot{m}_{\rm cool} = 4 \pi\, \rho_g(r_{\rm cool})\, r_{\rm cool}^2\, \dot{r}_{\rm cool}
\label{mdotcool}
\end{equation}
to determine the instantaneous cooling rate as
\begin{equation}
\dot{m}_{\rm cool} = \frac{1}{2}\, \left(\frac{r_{\rm cool}}{R_{\rm vir}}\right) \left(\frac{m_{\rm hot}}{t_{\rm cool}}\right) ~.
\label{coolstatic}
\end{equation}
Equation \ref{coolstatic} is valid as long as $r_{\rm cool} < R_{\rm vir}$, which marks when the system is in the hot-halo regime, as discussed above. Otherwise, infalling gas is in the cold-accretion regime, and we assume it deposits at the centre of the halo on a free-fall timescale, $R_{\rm vir} / V_{\rm vir}$.  In the presence of radio mode AGN heating, this cooling rate is modified with the addition of an inner heating radius.  We define our new hybrid cooling/heating model further in Section \ref{sec:radiomode}.

%% file: section07.tex
\section{The disk: cold gas, star formation, and metal enrichment}
\label{sec:sf}

Cooling gas that settles in the centre of a halo is assumed to conserve angular momentum and spin up to form a rotationally supported disk. It is from this disk of cold gas that stars form. In the \textsc{sage} model, star formation proceeds as in C06. A more sophisticated accounting of the formation of new stars, including the tracking of atomic and molecular hydrogen to fuel star formation, will be introduced in future work.

As discussed in C06, we assume that only cold gas above a critical surface density threshold can form stars. Following the work of \cite{Kauffmann1996}, and assuming the gas is evenly distributed across the disk (note that assuming an exponential disk profile simply results in a minor adjustment to the normalisation), we find a critical mass given by
\begin{equation}
m_{\rm crit} = 3.8 \times 10^{9}\, \left( \frac{V_{\rm vir}}{200\,{\rm km\,s^{-1}}} \right) \left( \frac{r_{\rm disk}}{10\,{\rm kpc}} \right)\, {\rm M_{\bigodot}} ~.
\label{m_crit}
\end{equation}
Here, the disk radius is defined as
\begin{equation}
	r_{\rm disk} = \frac{3}{\sqrt{2}}\lambda R_{\rm vir} ~,
	\label{r_disk}
\end{equation}
which is thrice the disk scale length proposed by \citet*{Mo1998}, based on the properties of the Milky Way \citep{vandenBergh2000}.  The spin parameter of the dark halo, $\lambda$ \citep{Bullock2001}, is taken directly from the $N$-body simulation. 

A star formation rate can now been calculated from a Kennicutt-Schmidt-type relation \citep{Kennicutt1998}:
\begin{equation}
\dot{m}_* = \alpha_{\rm{SF}} \, \frac{(m_{\rm cold} - m_{\rm crit})}{t_{\rm dyn,\rm disk}}~,
\label{sfr}
\end{equation}
where $m_{\rm cold}$ is the total mass of cold gas and $\alpha_{\rm{SF}}$ is the star formation efficiency. In other words, a fraction $\alpha_{\rm{SF}}$ of gas above the threshold is converted into stars in a disk dynamical time $t_{\rm dyn,{\rm disk}} = r_{\rm disk}/V_{\rm vir}$.

\begin{figure}
	\centering
	\includegraphics[width=0.45\textwidth]{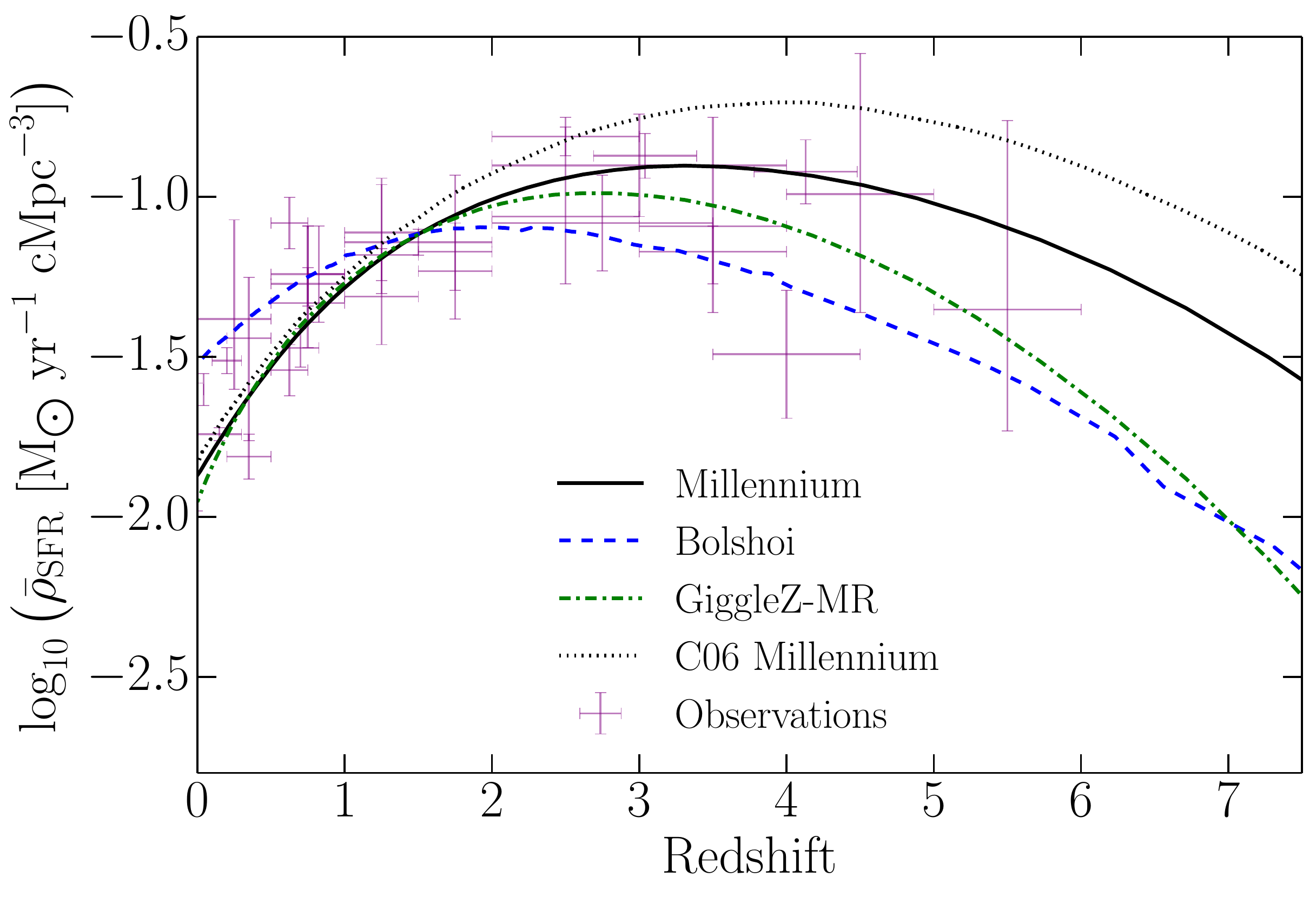}
	\caption{Average star formation rate density history produced by \textsc{sage} for each cosmological simulation compared to observations and the C06 model.  Observational data were originally compiled and corrected by \citet[][see their table A2 for a complete list of references]{Somerville2001}.}
	\label{fig:sfrd}
\end{figure}

This model of star formation, combined with the feedback processes described in the below sections, produces a galaxy population whose combined star formation rate density evolution is consistent with the observed Universe, as shown in Fig.~\ref{fig:sfrd}.  The supporting observational data were compiled by \citet{Somerville2001}, which we used as a constraint for \textsc{sage}.  Observational constraints on star formation rate density are non-trivial to obtain, naturally leading to large uncertainties \citep[for discussion, see][]{Somerville2001,Springel2003b}.  For example, in \textsc{sage} the time of peak star formation shows dependence on the $N$-body simulation: $z \sim 2$ for Bolshoi,  $z \sim 2.5$ for GiggleZ, and $z \sim 3$ for Millennium, yet they all agree with the broadness of the real data.  Millennium exhibits a systematically higher star formation rate than the other simulations for $z \ga 2$, which can be attributed to its larger $\sigma_8$ value.  Despite a largely identical star formation law to C06, variations in other parts of the model in \textsc{sage} (e.g.~AGN feedback) affect the star formation histories of galaxies noticeably.  As seen in Fig.~\ref{fig:sfrd}, the C06 model predicted star formation rates that were too high on average at high redshift.

\begin{figure}
	\centering
	\includegraphics[width=0.45\textwidth]{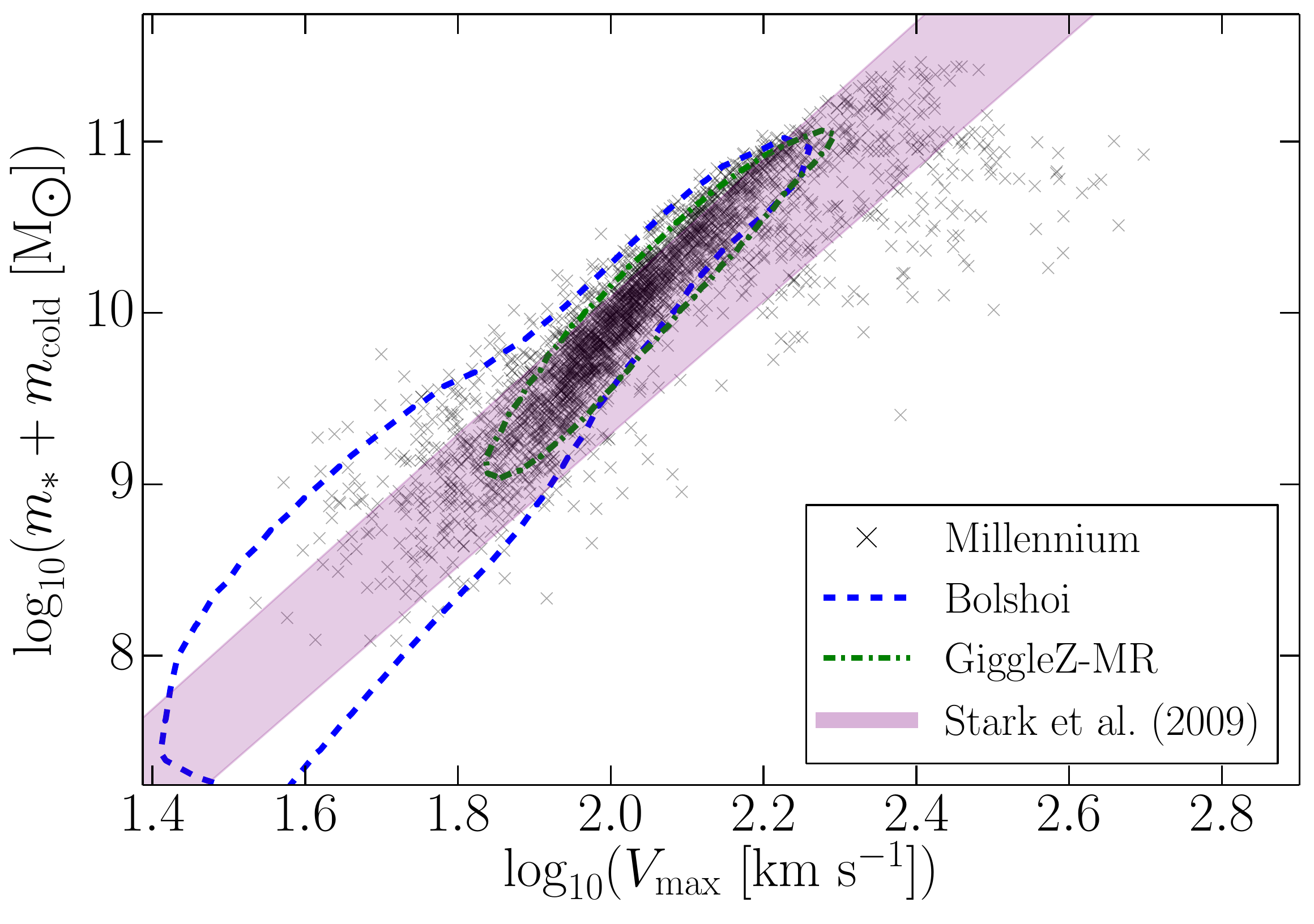}
	\caption{Baryonic Tully--Fisher relation for \textsc{sage} galaxies at $z=0$.  2500 representative galaxies (randomly sampled within the axes) with bulge-to-total ratios between 0.1 and 0.5 are plotted from Millennium, while contours encapsulating 68 percent of galaxies of the same cut are shown for the other $N$-body simulations.  The maximum Keplerian velocity for the subhalos, $V_{\rm max}$, represents the rotational velocity of galaxies.  Appropriately, only central galaxies were considered.  The thick strip represents the scatter in the observed relation \citep{Stark2009}.}
	\label{fig:btf}
\end{figure}

A well-known constraint on the properties of nearby disk galaxies is the correlation between luminosity (a proxy for stellar mass) and rotation velocity \citep{Tully1977}.  An even tighter correlation is found if one considers the contribution of cold gas to the mass of the systems: the so-called Baryonic Tully--Fisher relation \citep{McGaugh2000}.  For each of the $N$-body simulations, \textsc{sage} successfully produces Sb and Sc Hubble-type galaxies (proxied by a bulge-to-total ratio cut between 0.1 and 0.5) that closely match this relation, as shown in Fig.~\ref{fig:btf}.  The observational backdrop shows the fitted relation from \citet{Stark2009} with their random uncertainties; i.e.~we plot $\log_{10}\left((m_{*}+m_{\rm cold}) / \mathrm{M}_{\bigodot}\right) = (3.94 \pm 0.07)\log_{10}(V_{\max} / \mathrm{km~s}^{-1}) + (1.79 \pm 0.26) + 2\log_{10}(0.75/0.73)$, where the last term accounts for the assumed value of $h$.  We note that we have approximated the flat rotation velocity \citep[as used by][]{Stark2009} of \textsc{sage} galaxies by their halo's maximum value, $V_{\rm max}$.  We also note that the contours for Bolshoi show more galaxies at low $V_{\rm max}$ on account of the simulation's higher mass resolution.

As stellar populations evolve, they enrich the interstellar medium with elements heavier than hydrogen and helium. Such metal enrichment is modelled in \textsc{sage} by assuming a yield $Y$ of metals are returned for each solar mass of stars formed. We deposit these metals into the cold disk of the galaxy, but more complicated models may wish to explore direct metal enrichment of the hot halo gas and beyond \citep*[e.g.][]{Shattow2015}. Furthermore, a fraction $\mathcal{R}$ of the mass of newly formed stars is recycled immediately back to the cold gas disk: the so-called `instantaneous recycling approximation' \citep[see][]{Cole2000}. Our treatment of metals and recycling are identical to that described in C06 (and \citealt{deLucia2004} before), aside from differences in the assumed $Y$ and $\mathcal{R}$ parameters, which are guided by a change to the \citet{Chabrier2003} initial mass function.

\begin{figure}
	\centering
	\includegraphics[width=0.45\textwidth]{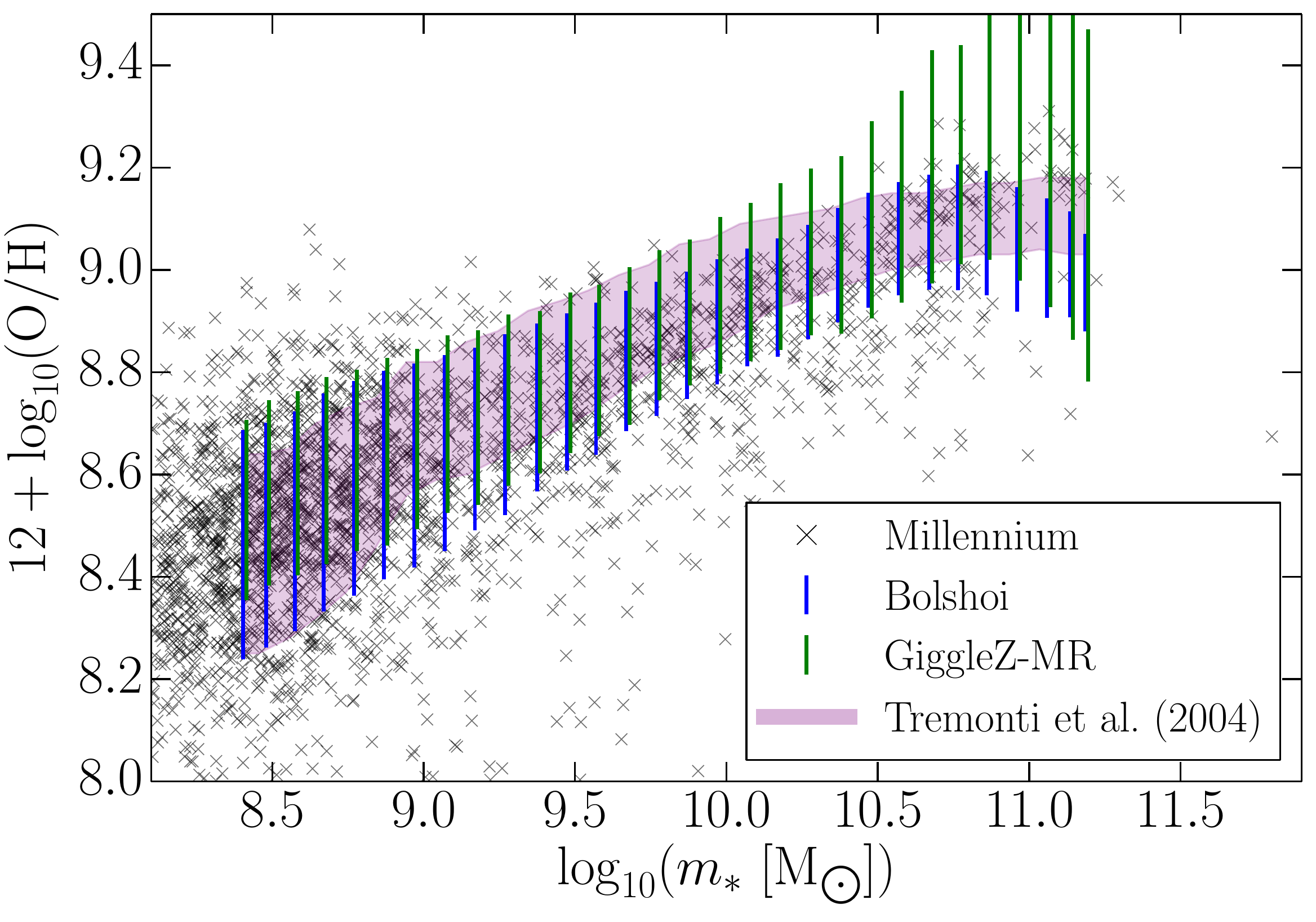}
	\caption{Stellar mass -- gas metallicity relationship for $z=0$ central galaxies in \textsc{sage}.  2500 representative galaxies are plotted in the axes from Millennium, while vertical lines cover the 16$^{\rm th}$--84$^{\rm th}$ percentile range for bins of stellar mass for the other $N$-body simulations.  The shaded region compares the 16$^{\rm th}$--84$^{\rm th}$ percentile range from observational data published by \citet{Tremonti2004}, binned identically to the Bolshoi and GiggleZ-MR data.}
	\label{fig:massmet}
\end{figure}

Through the measurement of optical nebular emission lines supplemented by galaxy photometry, \citet{Lequeux1979} were among the first to show evidence for a correlation between total mass and gas metallicity for galaxies.  Furthering this, \citet[][and see references therein]{Tremonti2004} showed there is a near quadratic relationship in logarithmic space for the oxygen abundance in galaxies, calculated based on the model of \citet{Charlot2001}, as a function of their stellar mass.  In Fig.~\ref{fig:massmet}, we show that the redshift-zero population of \textsc{sage} galaxies is representative of this observed relation for each $N$-body simulation by comparing against the 16$^{\rm th}$--84$^{\rm th}$ percentile range of \citet[][cf.~their table 3]{Tremonti2004}, adjusting for a conversion from a \citet{Kroupa2001} to \citet{Chabrier2003} initial mass function: $m_{*,\mathrm{Chabrier}} = (5/6) m_{*,\mathrm{Kroupa}}$.  Our choice for the value of the yield is almost entirely driven by this constraint.

The scatter seen in Fig.~\ref{fig:massmet} for the simulations is larger than the observational data on average (up to $\sim$75\%), although Bolshoi in particular shows strikingly good agreement.  The scatter for Millennium and GiggleZ-MR are almost identical across the full mass range, which can be seen in more detail by the individual Millennium points.

%% file: section08.tex
\section{Supernova feedback and the galactic fountain}
\label{sec:sn}

For each episode of new star formation, very massive stars have lifetimes much shorter than the typical time resolution of the simulations on which \textsc{sage} is built. Such stars end as supernovae. Supernovae play an important role in the life-cycle of a galaxy, injecting metals (as discussed above), mass and energy into the surrounding interstellar medium. Our modelling of supernova feedback follows that used in C06. 

First, we assume that supernova winds remove cold gas from the disk, which in turn acts to suppresses star formation. The rate at which disk gas is driven from the galaxy, $\dot{m}_{\rm reheated}$, is proportional to the rate at which new stars are forming:
\begin{equation}
\dot{m}_{\rm reheated} = \epsilon_{\rm{disk}}\, \dot{m}_*~.
\label{reheated}
\end{equation}
The proportionality constant, $\epsilon_{\rm{disk}}$, is typically referred to as a mass-loading factor.

More generally, the energy released by supernovae during the star formation episode can be approximated as
\begin{equation}
\dot{E}_{\rm SN} = \frac{1}{2}~\epsilon_{\rm{halo}}~\dot{m}_*\, V_{\rm SN}^2~,
\label{SN_energy}
\end{equation}
where $\frac{1}{2} V_{\rm SN}^2$ is the mean energy in supernova ejecta per unit mass of stars formed, $V_{\rm SN} = 630\,{\rm km\, s}^{-1}$ taking the commonly accepted value, and $\epsilon_{\rm{halo}}$ quantifies the efficiency with which this energy is able to reheat disk gas.
If this reheated disk gas were moved to the hot halo without changing its specific energy, the total thermal energy of the hot gas would change by
\begin{equation}
\dot{E}_{\rm hot} = \frac{1}{2}\, \dot{m}_{\rm reheated} V_{\rm vir}^2~.
\label{deltaEhot}
\end{equation}
Equations~\ref{SN_energy} and \ref{deltaEhot} allow us to determine the \emph{excess} energy in the hot gas after reheating: $\dot{E}_{\rm excess} = \dot{E}_{\rm SN} - \dot{E}_{\rm hot}$. When $\dot{E}_{\rm excess}\!<\!0$ supernovae has failed to transfer enough energy with the reheated disk gas to unbind any hot halo gas. However $\dot{E}_{\rm excess}\!>\!0$ signals that there is enough energy in the system to unbind some of the hot halo and maintain virial balance. Ejected hot gas is accounted for in the model by placing it in an external reservoir:
\begin{equation}
\dot{m}_{\rm ejected} = \frac{\dot{E}_{\rm excess}}{E_{\rm hot}} m_{\rm hot} = \ \left( \epsilon_{\rm halo} \frac{V_{\rm SN}^2}{V_{\rm vir}^2} - \epsilon_{\rm disk} \right) \dot{m}_*\, .
\label{ejected}
\end{equation}
Here, $E_{\rm hot}= \frac{1}{2} \, m_{\rm hot} V_{\rm vir}^2$ is the total thermal energy of the hot gas. No gas is ejected when the right-hand side of Equation \ref{ejected} is less than zero; this signals that $\dot{E}_{\rm excess}\!<\!0$, as mentioned above.

Equation~\ref{ejected} has many desirable properties. In low-mass halos with shallow potentials, the feedback can be very destructive, with the possibility that the entire disk and halo gas be expelled from the system if the feedback (i.e. star formation) is strong enough. Conversely, in halos with virial velocities above $200\,{\rm km\, s}^{-1}$, the potential well is sufficiently deep that no amount of feedback will remove hot gas. Such halos develop and maintain very stable hot atmospheres throughout their lives (barring mergers or other cataclysmic events).

\subsection{Reincorporation of ejected gas}

In a dynamically evolving universe, gas that is ejected may not stay ejected forever. In \textsc{sage}, we adopt a modified version of that used in C06 to determine ejected-gas reincorporation. Previously, C06 assumed that a fixed fraction of the ejected material returned to the hot halo over a halo dynamical time, and that this was true for all halos. As addressed by \citet{Mutch2013a}, this model was poorly constrained, where alterations to the reincorporation parameter could significantly alter the number of low-mass systems. In fitting \textsc{sage} for multiple simulations, we found that a better match to the data can be obtained when we allow the reincorporation rate to increase for the more massive halos, and limit it to zero for the very lowest-mass halos. We do this by assuming the mass of ejected gas reincorporated per time-step is
\begin{equation}
\dot{m}_{\rm reinc} = \left(\frac{V_{\rm vir}}{V_{\rm crit}} - 1\right) \, \frac{m_{\rm ejected}}{t_{\rm dyn}}~,
\label{reincorporated}
\end{equation}
where $t_{\rm dyn} = R_{\rm vir} / V_{\rm vir}$, and assuming $V_{\rm vir} > V_{\rm crit}$ ($\dot{m}_{\rm reinc}=0$ otherwise).  Here, $V_{\rm crit} = k_{\rm reinc} V_{\rm esc}$, where $V_{\rm esc} = V_{\rm SN} / \sqrt{2}$ is the critical halo virial velocity, above which the supernova wind velocity is sufficient for the gas to escape, and $k_{\rm reinc}$ parameterises how efficient this idealised process actually is (cf.~Table \ref{tab:params}). In effect, $V_{\rm crit}$ sets the velocity scale below which no gas can reincorporate back into a hot halo. Conversely, far above this scale, ejected gas reincorporates in a fraction of the dynamical time, meaning it quickly becomes available for future cooling and hence star formation.  

It has been suggested by \citet{Henriques2013} that using a virial-mass dependence rather than virial-velocity one for the reincorporation rate could be conducive to reproducing observations across multiple epochs, in particular the galaxy stellar mass function. Our initial exploration of this variation has not been quite as positive however \citep*[nor was it for][]{White2015}, and hence we do not adopt it here.  In truth, this is an area for improvement for both \textsc{sage} and other popular models. How to rectify the mass evolution of galaxies is an ongoing area of research deserving of far more investigation than the scope of this paper.

Note that the ejected gas component of our model need not be physically removed from the system. It merely marks gas that is unable to cool into the disk to form stars for a period of time. Such evacuated gas and metals may still play a part in a galaxy's evolution, simply one at a later epoch in its history.

%% file: section09.tex
\section{Supermassive black holes and their feedback}
\label{sec:agn}

Feedback from active supermassive black holes is now known to play a critical role regulating the life-cycle of many types of galaxies. C06 distilled a number of popular ideas about the interplay of active galactic nuclei (AGN) and galaxies into a simple picture of how this feedback shapes the properties of the low-redshift galaxy population \citep[see also][]{Bower2006,Cattaneo2006}. \cite{Kitzbichler2007} extended their comparison to higher redshift, specifically looking at galaxy masses, luminosities and number counts. Other authors have since adopted similar models and developed the basic framework to further explore how AGN evolve \citep[e.g.][]{Marulli2008,Somerville2008,Guo2011}.

C06 broke AGN into two general classes, loosely dubbed the `quasar mode' and `radio mode'. These two classes can be distinguished by their triggering mechanism, lifetime, and accretion rate. We maintain this two-mode approach in \textsc{sage} but have updated both, for which we describe each in turn below.

We note that we do not consider the AGN luminosity function in the calibration or results of the model.  This would require additional levels of modeling beyond the scope of this paper.  We do, however, plan to include more complex AGN physics to explore more observables in later developments of the model.

\subsection{The radio mode}
\label{sec:radiomode}
Radio mode feedback was introduced into semi-analytic models to solve the cooling flow problem, where the over-accretion of cooling gas onto the central galaxy led to massive galaxy properties that were inconsistent with observations (too massive, too blue, too disky). The first implementations were simple and framed either phenomenologically, which attempted to infer a black hole accretion rate (and hence feedback) based on the local black hole and gas properties (C06), or in terms of a sharp cutoff in cooling when the halo or galaxy evolved past a chosen critical state \citep{Bower2006}, e.g.~a mass threshold.

The model we employ in \textsc{sage} is an enhancement of the Bondi--Hoyle accretion model described in section~5.2 of C06, which is also used by \cite{Somerville2008}. In this model, hot gas accretes onto the central black hole at a rate approximated using the Bondi-Hoyle formula \citep{Bondi1952}:
\begin{equation}
\dot{m}_{\rm Bondi} = 2.5 \pi G^2\, \frac{m_{\rm BH}^2\, \rho_0}{c_{\rm s}^3}~. 
\label{physicalBH}
\end{equation}
Here, the sound speed, $c_{\rm s}$, is approximated by the virial velocity of the parent halo, $V_{\rm vir}$, while $\rho_0$ is the density of hot gas around the black hole. To find $\rho_0$, C06 equated the sound travel time across a shell of diameter twice the Bondi radius ($r_{\rm Bondi} \equiv 2G m_{\rm BH}/c_{\rm s}^2$) to the local cooling time (Equation~\ref{tcool}), the so-called `maximal cooling flow' model of \citet{Nulsen2000}. Inserting this density back into Equation~\ref{physicalBH} allows us to write the accretion rate as a function of local temperature and black hole mass alone:
\begin{equation}
\label{eq:radioacc}
\dot{m}_{\rm BH,R} = \kappa_{\rm R}\ \frac{15}{16} \pi G\ \bar{\mu} m_p\ \frac{kT}{\Lambda}\ m_{\rm BH}~.
\end{equation}
In a departure from C06, we insert a `radio mode efficiency' parameter, $\kappa_{\rm R}$, to the right-hand side of Equation \ref{eq:radioacc}.  While we have added this term by hand, it allows us to counteract the approximations used in the derivation of Equation \ref{eq:radioacc} and to modulate the strength of black hole accretion (and subsequently radio mode feedback) within \textsc{sage}.  As such, the ``best'' value need not be 1.  In the present work, we employ a default value of 0.08 (see below).

The accretion rate given by Equation~\ref{eq:radioacc} enables us to estimate the luminosity of the black hole in the radio mode, $L_{\rm BH,R} = \eta \,\dot{m}_{\rm BH,R} \,c^2$, where ${\eta = 0.1}$ is the standard\footnote{``Standard'' in terms of what has been adopted in the literature (e.g.~\citealt*{DiMatteo2005}; \citealt{Sijacki2007,Booth2009,Dubois2014c}; \citealt*{Dubois2014b}; \citealt{Henriques2014,Schaye2015}), a value which falls between the efficiency expected for a non-spinning and maximally spinning black hole (\citealt*{Bardeen1972}; also see fig.~1 of \citealt{Maio2013}).  Many authors cite \cite{Shakura1973} as the justification for $\eta = 0.1$, despite their using $\eta =0.06$.  Recent hydrodynamic works have started to use $\eta = 0.2$ instead \citep[e.g.][]{Hirschmann2014,Sijacki2015}, motivated by observations of more-luminous AGN \citep[][anti-respectively]{Yu2002,Davis2011}.} efficiency with which inertial mass is liberated upon approaching the event horizon, and $c$ the speed of light. We assume this luminosity acts as a source of heating that offsets the energy losses from the cooling gas. If enough energy from the central AGN is injected into the cooling flow, it can be turned off entirely, leading to longer-term quenching, which was the focus of the work by C06. If we define the specific energy of gas in the hot halo as $\frac{1}{2} V_{\rm vir}^2$, and heating as adding this amount of energy per unit mass to offset cooling, then the heating rate from radio mode feedback can be written as
\begin{equation}
\dot{m}_{\rm heat} = \frac{L_{\rm BH,R}}{\frac{1}{2} V_{\rm vir}^2}~.
\label{BHheating}
\end{equation}

\subsubsection{A more self-consistent treatment of the cooling--heating cycle}

In C06, the heating rate given by Equation~\ref{BHheating} was subtracted off the cooling rate given by Equation~\ref{coolstatic} to determine an effective AGN-adjusted cooling rate. However, despite its ability to reproduce many key local galaxy properties, there remained important limitations. Significantly, in the C06 model (and many similar radio mode models), cooling and heating are decoupled, meaning the latter only offsets the former after both have been independently calculated. In the real Universe, one would expect episodes of AGN activity to have a lasting (or, at least, extended) effect on the surrounding gas, modifying its temperature and density profile in such a way that would alter the later cooling. A coupled cooling--heating model is clearly a desirable refinement if one wants to make realistic predictions for e.g.~the X-ray halo cooling luminosity or AGN jet power. Unfortunately, there is no natural way to achieve such cooling--heating coupling within the paradigm of current semi-analytic models. 

To achieve a coarsely equivalent behaviour, we implement a simple idea in \textsc{sage} to create an updated AGN model. We assume that cooling gas is heated by radio mode feedback out to a particular radius, called $r_{\rm heat}$, interior to which the gas retains the memory of its past heating. This gas will \emph{never} cool thereafter. To determine $r_{\rm heat}$ we find the radius at which the energy deposited into the gas due to the radio mode equals the energy the halo gas interior to $r_{\rm heat}$ would lose if it were to cool onto the galaxy disk from Equation~\ref{coolstatic} alone. This is given by
\begin{equation}
r_{\rm heat} = \frac{\dot{E}_{\rm heat}}{\dot{E}_{\rm cool}}\, r_{\rm cool}~,
\label{rheat}
\end{equation}
where
\begin{equation}
\dot{E}_{\rm heat}= \frac{1}{2} \dot{m}_{\rm heat} V_{\rm vir}^2 \ \ {\rm and}\ \ \dot{E}_{\rm cool} = \frac{1}{2} \dot{m}_{\rm cool} V_{\rm vir}^2 ~,
\label{Eheatingcooling}
\end{equation}
and $\dot{m}_{\rm cool}$ and $\dot{m}_{\rm heat}$ are determined from Equations~\ref{coolstatic} and \ref{BHheating} above. The cooling rate given in Section~\ref{sec:cooling} is then modified by the presence of this heating radius. The new cooling rate, in the presence of current and past radio mode episodes, now becomes
\begin{equation}
\dot{m}'_{\rm cool} = \left(1 - \frac{r_{\rm heat}}{r_{\rm cool}}\right) \dot{m}_{\rm cool}~.
\label{newcoolstatic}
\end{equation}
This model assumes that only the gas between $r_{\rm heat}$ and $r_{\rm cool}$ can cool, and when $r_{\rm heat} > r_{\rm cool}$, cooling is effectively quenched (i.e. $\dot{m}'_{\rm cool} = 0$). To retain the memory of past heating episodes, the heating radius is never allowed to move inwards, only outwards.

\begin{figure}
	\centering
	\includegraphics[width=0.45\textwidth]{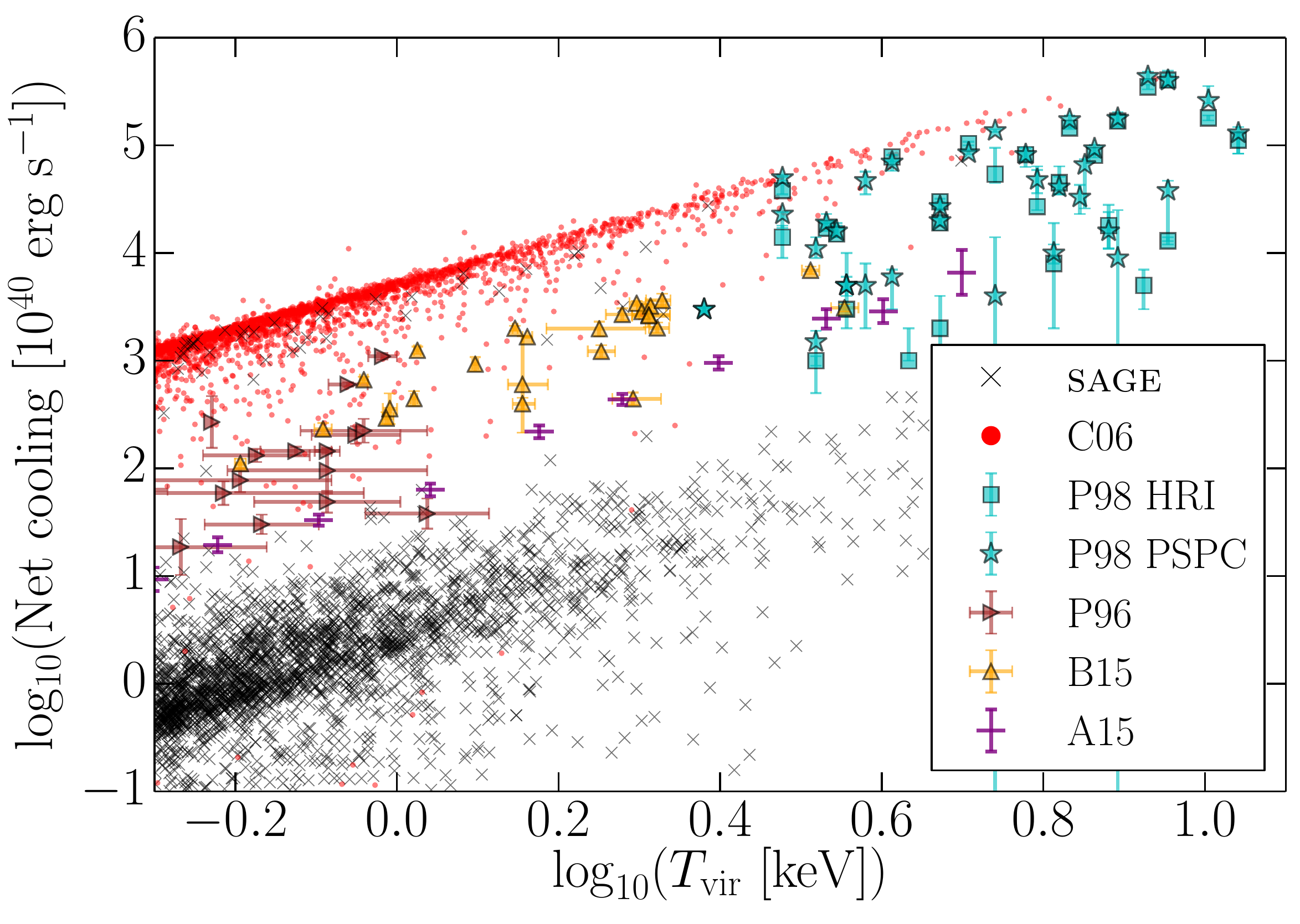}
	\caption{Cooling luminosity of hot gas in halos as a function of virial temperature at $z=0$.  Here we compare what the C06 model for cooling would have returned against \textsc{sage} for Millennium halos.  2500 randomly points are plotted within the axes for both the \textsc{sage} and C06 cooling methods.  Squares and starred points with error bars compare observational data of galaxy clusters from \citet{Peres1998}, measured with the High Resolution Imager (HRI) and Position Sensitive Proportional Counter (PSPC) at \emph{ROSAT}, respectively.  Triangular points with errors show observational data of galaxy groups \citep{Ponman1996,Bharadwaj2015}.  Stacked X-ray observations surrounding locally brightest galaxies from \citet{Anderson2015} are given by the horizontal marks with errors.}
	\label{fig:coolinglum}
\end{figure}

This new model behaves almost identically to that used in C06, but with the added benefit that a more realistic cooling rate can be extracted from the cooling--heating cycle rather than simply an upper limit.  To demonstrate this point, we show the cooling rates predicted by \textsc{sage} (i.e.~Equation \ref{newcoolstatic}) against those from the C06 model (i.e.~Equation \ref{coolstatic} less Equation \ref{BHheating}) for the same halos in Fig.~\ref{fig:coolinglum} at $z=0$.  The cooling rates are shown against halo virial temperature (Equation \ref{eq:Tvir}).  For comparison, also plotted are various X-ray observations of hot gas halos surrounding galaxy clusters and groups (\citealt{Ponman1996} (P96); \citealt{Peres1998} (P98); \citealt{Anderson2015} (A15); \citealt{Bharadwaj2015} (B15)), as marked.\footnote{We have used a subset of the observational data in the literature to cover the plotted range of virial temperatures.  Many other data exist (see, e.g., fig.~7 of \citealt{Anderson2015} and references therein) and they are all consistent with the conclusions we draw here.}  Note that P98 measure the cooling luminosity (X-ray luminosity inside the cooling radius), whereas P96 and B15 measure the bolometric luminosity (X-ray luminosity inside $R_{500}$), for which the cooling luminosity of such systems will be lower. Furthermore, the observations statistically favour the brightest systems at any given mass scale, while the model draws more broadly from the wider galaxy and halo population.  The \citet{Anderson2015} data are stacked observations of X-ray emission around ``locally brightest galaxies''.  These data should, in principle, be more representative of an average halo (suffering optical selection bias rather than X-ray selection bias), and therefore more directly comparable to \textsc{sage}.  We have plotted the maximum of their measurement and bootstrapping errors for each binned point.

The upper-limit nature of the C06 model is highlighted in Fig.~\ref{fig:coolinglum} by the tight band of points between $10^{0.4} < T_{\rm vir} / \mathrm{keV} < 10^{0.6}$ relative to the observations. In this model, the radio feedback required to offset the excessive cooling demanded $\kappa_R = 1$, with correspondingly higher absolute heating rates.  With less cooling to counteract, our nominal value for $\kappa_R$ has been reduced to 0.08 for \textsc{sage}. In other words, our required heating rates are significantly lower than before, the results of which can be more meaningfully compared with observations.

In truth, if an AGN has been off for an extended period of time one might expect the hot gas in the halo to return to its previous state and cool at closer to the maximal rate.  Ideally, $r_{\rm heat}$ would shrink during this time, which is most likely why our cooling rates lie below the observations.  In this sense, our prevention of $r_{\rm heat}$ moving back inwards places a lower limit on the cooling rates of halos. The addition of $r_{\rm heat}$ in \textsc{sage} is still a step in the right direction though, and goes some way to include often missed effects, such as the observed entropy floors in many cluster hot gas profiles.  We leave a more thorough treatment of the cooling--heating cycle for a future study.

\subsection{The quasar mode}
\label{sec:quasarmode}

In most simulations of galaxy formation, quasars are triggered by mergers or from some form of instability in the disk. The key requirement is to funnel gas into the galactic centre on a very short time-scale, which results in high black hole accretion rates and rapid black hole growth. Hence, the quasar mode is the dominant mode through which a black hole gets its mass.

Galaxy mergers are described below in Section~\ref{sec:merging}. Of relevance here, to model the effect of mergers on black hole growth we follow the work of \cite{Kauffmann2000}, as well as the enhancements in C06, and assume that mergers trigger black hole growth according to the phenomenological relation
\begin{equation}
\Delta m_{\rm BH,Q} = \frac{f'_{\rm BH} \ m_{\rm cold}}{1 +  (280\,{\rm{km\,s^{-1}}}/V_{\rm vir})^2}~, 
\label{accretionQ}
\end{equation}
where
\begin{equation}
f'_{\rm BH} = f_{\rm BH}\ (m_{\rm sat}/m_{\rm central})
\label{fBHparameter}
\end{equation}
is an accretion efficiency parameter with constant $f_{\rm BH}$, which controls the fraction of cold gas accreted by a black hole and is modulated by the satellite-to-central galaxy merger mass ratio.

Disk instabilities can similarly lead to rapid black hole growth. In Section~\ref{sec:instability} we describe our instability implementation, which is similar to that used in C06. For the sake of instability-driven accretion here, we modify Equation~\ref{accretionQ} by taking $f'_{\rm BH} = f_{\rm BH}$ and substitute the mass of unstable cold gas for $m_{\rm cold}$.

Although C06 used the quasar mode to grow black holes, they did not include quasar feedback when accounting for the evolution of the surrounding baryons. In \textsc{sage}, quasar mode feedback is included. This mode has little effect on the local galaxy population but can have a significant impact on early universe galaxy formation. An exploration of quasar mode feedback is left for a future paper; here we simply describe its implementation.

In the absence of a detailed understanding of how quasar accretion and feedback operates \citep*[but see][and references therein]{Lynden-Bell1969,Novikov1973,Costa2014,Stevens2015}, we adopt a simple phenomenological model that is consistent with the quasar mode narrative. When a merger or disk instability occurs and the black hole has undergone some form of rapid accretion, we assume a quasar wind follows with luminosity $L_{\rm BH,Q} = \eta \,\dot{m}_{\rm BH, Q} \,c^2$, where ${\eta = 0.1}$ as before. This is used to calculate the total energy contained in the quasar wind,
\begin{equation}
E_{\rm BH,Q} = \kappa_{\rm Q}\, \frac{1}{2}\, \eta\, \Delta m_{\rm BH,Q}\, c^2 ~, 
\label{Qfeedback}
\end{equation}
where $\kappa_{\rm Q}$ parametrises the efficiency with which the wind influences the surrounding gas as it escapes the galaxy and halo.
Next, we calculate the total thermal energies in both the cold disk gas and hot halo gas:
\begin{equation}
E_{\rm cold} = \frac{1}{2}\, m_{\rm cold}\, V_{\rm vir}^2 \ \ {\rm and}\ \ E_{\rm hot} = \frac{1}{2}\, m_{\rm hot}\, V_{\rm vir}^2 ~.
\label{Ecoldhot}
\end{equation}
Simply put, if the total energy in the quasar wind (Equation~\ref{Qfeedback}) exceeds the total energy in the cold disk gas we blow out the cold gas and associated metals to the ejected gas reservoir. If the quasar energy is greater than the combined total energy in the cold gas and hot halo gas, the quasar wind instead ejects both the cold and hot gas (and metals) from the halo. This is an ``all or nothing'' approach that is ripe for development.

\subsubsection{Black hole population}

\begin{figure}
	\centering
	\includegraphics[width=0.45\textwidth]{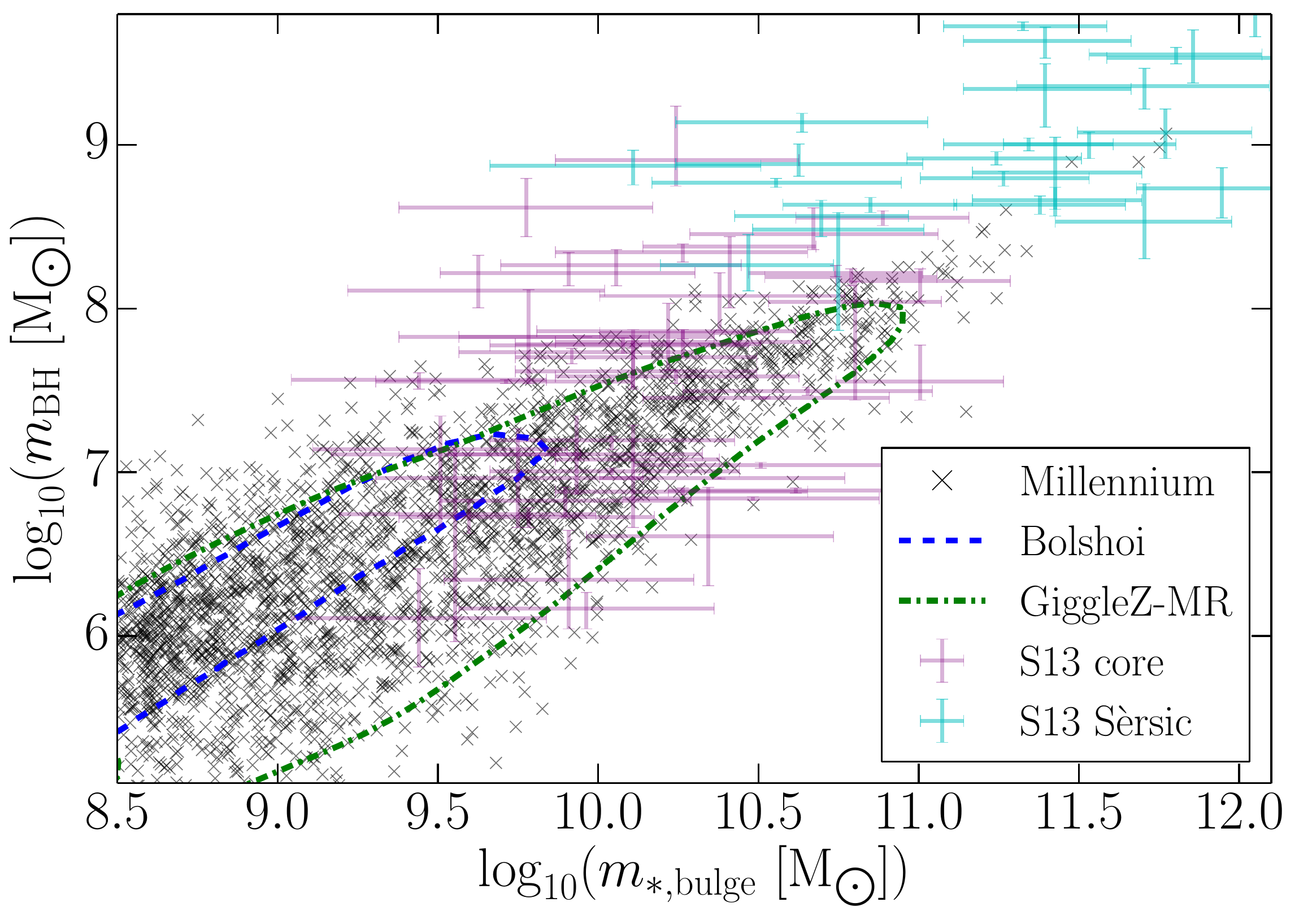}
	\caption{Black hole--bulge mass relation for \textsc{sage} galaxies.  2500 representative galaxies are plotted for Millennium within the axes, while contours encapsulating 68 percent of systems containing black holes and bulges are shown for the other $N$-body simulations..  Error bars compare observational data from \citet{Scott2013}, where cyan represents S\`{e}rsic galaxies, and purple represents core-S\`{e}rsic.}
	\label{fig:bhbulge}
\end{figure}

While the radio mode regulates cooling in \textsc{sage}, the quasar mode is the dominant channel for black hole growth.  In Fig.~\ref{fig:bhbulge}, we show that our treatment of black hole evolution is in general agreement with the observed black hole--bulge mass relation.  We compare \textsc{sage} galaxies to the observed sample published in \citet{Scott2013}, which considers S\`{e}rsic and core-S\`{e}rsic galaxies (the latter with typically more-massive bulges) separately \citep[also see][]{Graham2015}.  Once again, observational statistics favour large bulge masses, while numbers are naturally greater for low-bulge-mass systems from the theory perspective.  The region of overlap spans over 1 dex in width though, and shows clear agreement.

We note that we do not consider either the quasar or radio AGN luminosity functions in the current work.  This would require additional layers of modelling that are beyond the scope of this paper.  We do, however, plan to include more complex AGN physics in later developments of \textsc{sage} that will allow us to explore more observables.

%% file: section10.tex
\section{Galaxy mergers and intra-cluster stars}
\label{sec:merging}
The new \textsc{sage} model treats satellite galaxies somewhat differently to the previous C06 model. In C06, satellites had their hot halo instantly stripped upon infall, and their orbits were followed using the host subhalo position until the subhalo dark matter mass stripped below the resolution limit of the simulation. Upon losing the subhalo, an analytic merger time was calculated assuming the dynamical friction model of \cite{Binney1987} and the properties of the subhalo in the time-step before it was lost:
\begin{equation}
t_{\rm friction} = 1.17 \frac{V_{\rm vir} R_{\rm vir}^2}{G
  m_{\rm sat} \ln \Lambda} ~.
\label{merging_time}
\end{equation}
Here, $m_{\rm sat}$ is the total mass of the subhalo/satellite system (dark matter plus baryonic), and the Coulomb logarithm $\ln \Lambda$ is approximated by $\ln (1+M_{\rm vir}/m_{\rm sat})$. The `orphan' (i.e. subhalo-less) galaxy was then allowed to survive until this clock ran out, after which it was merged with the central galaxy.

A few important consequences resulted from this satellite galaxy model. In particular, it was realised that the colours of satellite galaxies were too red, as noted by many authors \citep[C06;][]{Weinmann2006,Guo2011}. This was because the instantaneous stripping of hot gas was imposing an artificial quenching mechanism on satellites, leading to premature suppression of star formation \citep[also see][]{Font2008}. Furthermore, in the C06 model, all satellites were assumed to merge with the central galaxy once their dynamical friction clock reached zero, whereas we know from observations that many satellites are shredded to pieces well before merging, and instead instead become part of an intra-group or intra-cluster stellar halo.  While scatter in a mass-dependent dynamical friction formula is inevitable, hydrodynamic simulations have since suggested there is a better generic fit than Equation \ref{merging_time} (\citealt{Jiang2008}; \citealt*{Jiang2010}).

In \textsc{sage}, we allow for the evolution of the satellite population by treating them more like central galaxies. Hot-halo stripping now happens in proportion to the dark matter subhalo stripping, rather than instantaneously. Any hot gas present in the subhalo is allowed to cool onto the satellite in the usual way. Upon infall, a merger time is calculated for the satellite, using the same dynamical friction formula as above (Equation \ref{merging_time}). We take this as the \emph{average} merger time expected for systems of similar properties. We then follow the satellite with time and measure the ratio of subhalo-to-baryonic mass. When this ratio falls below a critical threshold, $f_{\rm friction}$ (typically 1: cf.~Table \ref{tab:params}), we compare its current survival time with the average time determined at infall. If the subhalo has survived longer than average, then we say that the subhalo/satellite system was more bound than average and merge it with the central in the standard way. On the other hand, if the subhalo/satellite mass ratio has fallen below the threshold sooner than average, then we argue that the system was instead loosely bound and more susceptible to disruption. In this case, we add the satellite stars to a new intra-cluster stars component, and any remaining gas goes to the parent hot halo.  This omission of orphan galaxies is one of \textsc{sage}'s notable points of difference to other semi-analytic models currently in the literature.

Once the occurrence of a galaxy--galaxy merger has been identified, we check the satellite-to-central baryonic mass ratio. If the ratio is above a threshold $f_{\rm major}$ -- in \textsc{sage} set at a default of 0.3 -- we say the merger is `major'. In a major merger the disks of both galaxies are destroyed and all stars are combined to form a spheroid. Otherwise the merger is `minor', and only the satellite stars are added to the central galaxy bulge. Furthermore, any cold gas present in either system can lead to a starburst, as described below in Section~\ref{sec:bursts}.

\begin{figure}
	\centering
	\includegraphics[width=0.45\textwidth]{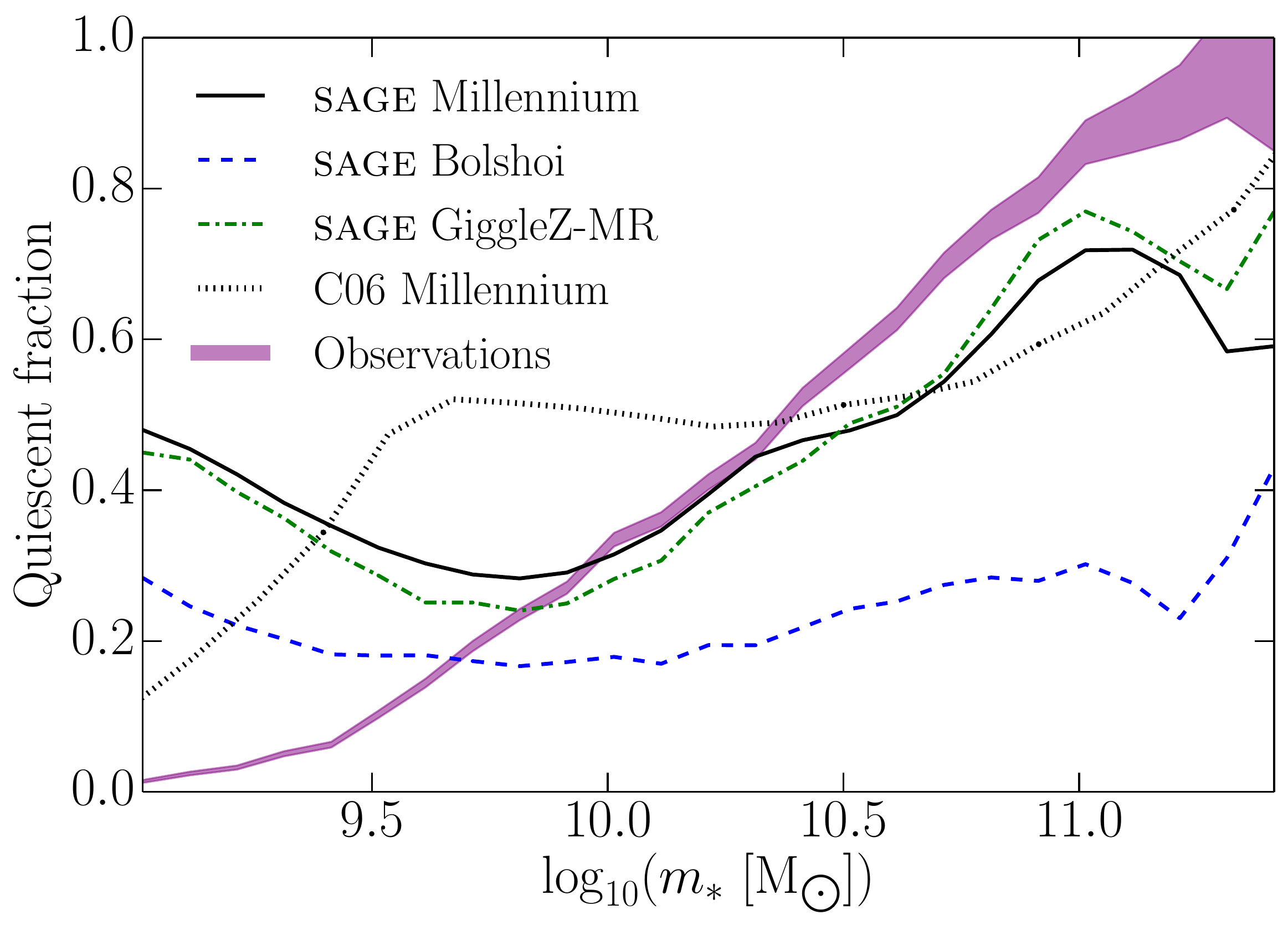}
	\caption{Fraction of quiescent (i.e.~low specific star formation rate) satellites at $z=0$ in \textsc{sage} as a function of stellar mass.  Compared are observed galaxies at $z<0.05$ from SDSS along with the C06 model.  Too many quiescent galaxies are consistently seen in the models for $m_{*} \lesssim 10^{10}\, \mathrm{M}_{\bigodot}$.}
	\label{fig:quiescent}
\end{figure}

\textsc{sage} still overproduces the fraction of quiescent satellite galaxies, a problem shared with other semi-analytic models \citep[e.g.][]{Guo2011}. In fact, this is true for galaxies at lower masses ($m_* \lesssim 10^{10}\, \mathrm{M}_{\bigodot}$) in general.  In Fig.~\ref{fig:quiescent}, we show the fraction of quiescent\footnote{It is worth noting that a red fraction, defined by a cut in colour (typically $g-r$), is not equivalent to a quiescent fraction, defined by a cut in specific star formation rate.} galaxies from each $N$-body simulation, determined as those with specific star formation rates $<10^{-11}\, \mathrm{yr}^{-1}$, as a function of stellar mass.  To compare to real galaxies, we calculate a quiescent fraction (by the same definition as the model galaxies) from Sloan Digital Sky Survey \citep[SDSS;][]{York2000} galaxies, specifically Data Release 7 \citep{Abazajian2009}.  Stellar masses and star formation rates come from the MPA-JHU catalogue,\footnote{http://wwwmpa.mpa-garching.mpg.de/SDSS/DR7/} where star formation rates are based on \citet{Brinchmann2004}.  We bin these data in stellar mass of width 0.1 dex and display Poisson errors for the quiescent fraction.  We only include galaxies at $z < 0.05$.  Also compared against these data is the quiescent fraction for C06 galaxies.  While \textsc{sage} has improved the quiescent fraction at masses around $10^{10}\, \mathrm{M}_{\bigodot}$, the problem at lower masses, i.e.~in satellites, persists (as does an overproduction of star-forming galaxies at high masses).  We note that because Millennium was our primary simulation for constraining the model, and GiggleZ-MR is of similar resolution and used the same halo finder, these simulations produce more promising results than Bolshoi.  Due to a number of effects, including Bolshoi being more affected by reionization, baryons cool in Bolshoi halos at a later time, therefore systematically raising the specific star formation rates of galaxies at $z=0$ (also seen in Fig.~\ref{fig:sfrd}).

We acknowledge that removing orphans may have consequences for galaxy clustering \citep{Saro2008,Guo2011}, which we intend to address in a follow-up study.  As it stands, however, most cosmological simulations have sufficiently high subhalo resolution to reproduce the observed clustering of the galaxy population through the halo occupation distribution model \citep[see][]{Berlind2002}, which also lacks an orphan population.  An orphan-less semi-analytic model is significantly more modular and transportable between differing simulations, which is a desirable property in the \textsc{sage} codebase.

%% file: section11.tex
\section{Disk instabilities}
\label{sec:instability}

Galaxies also transform through instabilities that occur within a galaxy disk itself. Again, we follow C06: after each episode of star formation, we determine a stability criterion \citep{Mo1998} using the disk mass, $m_{\mathrm{disk}}$; radius, $r_{\mathrm{disk}}$; and circular velocity, $V_c$, approximated by the maximum circular velocity of the halo. For the disk to be stable the following inequality must be met:
\begin{equation}
\frac{V_c}{(G\, m_{\mathrm{disk}} / r_{\mathrm{disk}})^{1/2}} \ge 1 ~.
\label{disk_stability}
\end{equation}
When the left side of Equation~\ref{disk_stability} is less than unity, we transfer (in proportion) enough stellar and cold gas mass to the bulge to return the disk to stability. Unstable cold gas can both grow the central black hole (as described in Section~\ref{sec:quasarmode}) and lead to the formation of new stars in a starburst (as described in the next Section).

%% file: section12.tex
\section{Starbursts}
\label{sec:bursts}

Starbursts mark the rapid formation of new stars, triggered as a result of a specific event. This is opposed to the more typical quiescent star formation that goes on in the galactic disk and described in Section~\ref{sec:sf}. In \textsc{sage}, both galaxy mergers and disk instabilities act as triggers for starbursts.
\begin{itemize}
\item[] \noindent \textbf{Mergers:} When a galaxy merger has been identified, the resulting starburst occurs in proportion to the total sum of the cold gas in the two merging galaxies. \textsc{sage} treats starbursts using the implementation of \cite{Somerville2001}, where the fraction of cold gas converted to stars is given by
\begin{equation}
e_{\rm burst} = \beta_{\rm burst} (m_{\rm sat} / m_{\rm
  central})^{\alpha_{\rm burst}} ~.
\label{e_burst}
\end{equation}
The two parameters above are fixed at $\alpha_{\rm burst}=0.7$ and $\beta_{\rm burst}=0.56$, which provides a good fit to the numerical results of \cite{Cox2004} and also \cite{Mihos1994, Mihos1996} for merger mass ratios ranging from 1:10 to 1:1. For minor mergers, the new stars are added to the galactic bulge and the stability of the disk is subsequently checked.  As mentioned in Section \ref{sec:merging}, for major mergers, \emph{all} stars go to the spheroid, including those newly formed.\\

\item[] \noindent \textbf{Instabilities:} For instability-driven starbursts, $e_{\rm burst}$ is taken as the fraction of cold disk gas that is unstable (Section~\ref{sec:instability}), minus any gas that is accreted onto the central black hole (Section~\ref{sec:quasarmode}). All newly formed stars from a disk instability burst are then added to the bulge (adding them to the disk would simply leave the disk unstable).
\end{itemize}

Our starburst implementation for mergers is in contrast to the recent semi-analytic work of \cite{Padilla2014}; rather than applying Equation \ref{e_burst}, those authors instead check for an instability to drive a starburst after merging, by virtue of evolving the size of disks rather than assuming Equation \ref{r_disk}.

%% file: section14.tex
\section{Discussion and Summary}
\label{sec:discussion}

We have presented our new, publicly available semi-analytic model of galaxy formation and evolution, \textsc{sage}.  The model is based on that of \citet{Croton2006}, but has updated many of the transfer processes between baryonic reservoirs, including gas infall, cooling, heating, and reincorporation.  It further includes previously omitted quasar mode feedback for supermassive black holes and an intra-cluster star component for central galaxies.  In addition, \textsc{sage} has a unique take on satellite galaxies, whereby they are assumed to be disrupted or merge when they lose their dark matter subhalo, rather than surviving as orphans.  

The codebase that describes \textsc{sage} is modular and well suited to run on the halo merger trees of any cosmologically representative $N$-body simulation. In this paper we have shown the performance of \textsc{sage} on each of the Millennium, Bolshoi, and GiggleZ simulations using a common set of default parameters. With these, \textsc{sage} can successfully reproduce many observational properties of the local galaxy population simultaneously. Our calibrations include the redshift-zero stellar mass function, baryonic Tully--Fisher relation, stellar mass--gas metallicity relationship, black hole--bulge mass relation, and the average star formation rate density history. We find only minor differences between simulations in the predicted scaling relations. This insensitivity to halo finding and merger tree construction reflects the robust nature of the included physical prescriptions.

However, no model of the galaxy population can ever be considered complete. The physics that governs galaxy evolution can always be examined at a higher level of complexity, and in the current era of survey science our instruments continue to produce increasingly vast and rich data sets. Within these data, the properties of galaxies are being measured in increasingly refined detail. To stay current, \textsc{sage} is ripe for further development in a number of key areas:

\begin{itemize}
	\item More detailed modelling of gas outside of and within the halo and galaxy, including radio predictions for the large-scale distribution of atomic hydrogen, new baryonic reservoirs such as the circumgalactic medium and warm intergalactic medium, and the different phases of gas in the galactic disk, specifically its neutral and molecular hydrogen content.
	\item Improved modelling of the various stellar growth channels of disks and bulges in galaxies, their size and evolution, and how this modifies their broader predicted observed properties.
	\item An improved understanding of gas and star formation in satellite galaxies; in particular, the formation and abundance of low mass galaxies at high redshift, and the discrepancies found by many models with the quiescent fraction of satellites at low redshift.
	\item Expanding the model to produce predictions for the AGN population, such as radio jet luminosities and sizes, and the abundance of radio AGN as a function of host galaxy mass and redshift.
\end{itemize}

Orthogonal to its use to explore questions of science, the philosophy behind \textsc{sage} is one of transparency and reproducibility of scientific results. By making \textsc{sage} an open and community project, we are hoping to (a) widen the accessibility of such models to more astronomers, especially students; (b) enable wider development of the science modules, which will hopefully be useful to astronomers with more specific interests; and (c) increase the scrutiny of how such models are produced, their uncertainty and limitations, and their use in the literature. In this sense, \textsc{sage} is following a similar path to that already established by scientific codes such as \textsc{gadget-2} \citep{Springel2005Gadget2} and \textsc{galacticus}\footnote{https://sites.google.com/site/galacticusmodel/} \citep{Benson2012}, as well as many others (see the Astrophysics Source Code Library\footnote{http://ascl.net/} for further examples).

The default catalogues produced by \textsc{sage} and used here are publicly available for download at Swinburne University's Theoretical Astrophysical Observatory \citep[TAO,][]{Bernyk2014}. With TAO, users can additionally add a wide range of observational filters to produce apparent and absolute magnitudes, build custom light cones to mimic popular surveys, and create custom images of a mock galaxy population. These datasets form a solid basis for comparisons with survey data, although users may wish to locally re-run them using \textsc{sage} and tweak the parameters to further refine the model. When new simulations of significance become publicly available, and new models are build on top of them, we expect to use TAO to distribute these as well. 

Having an array of simulations and models on-hand, and being able to generate new ones as needed, will enable astronomers to explore the theoretical uncertainty between different mock datasets. This is especially important for models which claim to follow a similar underlying physical narrative, but with different technical implementation. Expanding the pallet of theoretical predictions available to observers will add valuable context when using such models to compare with and interpret observational results.